\documentclass[reprint,twocolumn]{revtex4-2}
\usepackage{amsfonts}
\usepackage{amsmath}
\usepackage{amssymb}
\usepackage{charter}
\usepackage{graphicx}
\usepackage{float}
\usepackage{amsmath}
\usepackage{amssymb}
\usepackage{charter}
\usepackage{graphicx}
\usepackage{subfigure}
\usepackage{graphicx}
\usepackage{epstopdf}
\usepackage{color}
\usepackage{tocvsec2}
\usepackage{enumerate}
\usepackage{graphicx}
\usepackage{dcolumn}
\usepackage{bm}

\begin{document}

\title{Enhancing Phase Estimation in a Hybrid Interferometer via Kerr
Nonlinearity and Photon Subtraction}
\author{Lifen Guo$^{1}$, Qingqian Kang$^{1,2}$, Zekun Zhao$^{1}$, Jifeng Sun$%
^{3}$, Teng Zhao$^{1}$, Cunjin Liu$^{1}$, Xin Su$^{1}$, Liyun Hu$^{1,\ast }$}
\affiliation{$^{{\small 1}}$\textit{Center for Quantum Science and Technology, Jiangxi
Normal University, Nanchang 330022, China}\\
$^{{\small 2}}$\textit{Department of Physics, Jiangxi Normal University
Science and Technology College, Nanchang 330022, China}\\
$^{{\small 3}}$\textit{School of Electronic and Information Engineering,
Nanchang Institute of Technology, Nanchang 330044, China}}
\thanks{hlyun@jxnu.edu.cn}

\begin{abstract}
We propose a high-precision phase estimation scheme in a hybrid
interferometer by synergistically combining a Kerr nonlinear phase shifter
and multi-photon subtraction operations. Using a coherent state and a vacuum
state as input resources, we systematically evaluate the phase sensitivity
via homodyne detection and analyze the quantum Fisher information as well as
the quantum Cram\'{e}r-Rao bound under both ideal and lossy conditions. Our
results show that the joint integration of Kerr nonlinearity and
multi-photon subtraction yields remarkable advantages over either technique
used alone. The proposed scheme enables the phase sensitivity to surpass the
standard quantum limit, exceed the conventional Heisenberg scaling ($1/N$),
and approach the super-Heisenberg scaling ($1/N^{2}$)-a direct consequence
of Kerr nonlinearity. More precisely, the super-Heisenberg scaling $\propto $
$1/N^{2}$ is the ultimate precision limit permitted by the $k=2$ Kerr
nonlinearity and does not violate the fundamental Heisenberg limit for
linear phase accumulation. Even under moderate internal photon loss, the
system maintains high precision and exhibits enhanced robustness to
decoherence. The Kerr nonlinearity introduces an intensity-dependent phase
shift proportional to the squared photon number, while multi-photon
subtraction tailors non-Gaussian states to strengthen phase information
extraction. Compared with existing schemes based on hybrid interferometers
or SU(1,1) interferometers, our architecture achieves superior precision and
stronger loss resilience. All components are experimentally accessible with
current quantum optical technologies. This work provides a promising route
for practical high-precision quantum metrology and quantum sensing.

\end{abstract}
\maketitle

\section{Introduction}

Precision phase measurement is a cornerstone of high-precision metrology and
has attracted considerable attention in recent years. By harnessing quantum
resources such as quantum entanglement and squeezing, it enables precise
estimation of unknown parameters, with critical applications in quantum
illumination \cite{1,2,3}, optical imaging \cite{4,5}, and quantum sensing
\cite{6}. The ultimate precision of any estimation is fundamentally bounded
by the quantum Cram\'{e}r-Rao bound (QCRB) \cite{7,8}, which provides a
universal lower limit independent of the measurement scheme.

Interferometry is a principal tool for phase estimation. Conventional
architectures include the Mach--Zehnder interferometer (MZI) \cite{9} and
the SU(1,1) interferometer (SUI) \cite{10,11}. The SUI replaces the passive
beam splitters (BSs) in the MZI with active optical parametric amplifiers
(OPAs) \cite{12,13}, making it a promising candidate for high-precision
parameter estimation. However, the SUI cannot fully exploit photons carrying
phase information because the second OPA absorbs some internal photons. To
overcome this drawback, the hybrid interferometer (HI) has been proposed, in
which the second OPA is replaced by a conventional beam splitter (BS). For
instance, Kong et al. \cite{14} demonstrated that the HI can achieve
signal-to-noise-ratio enhancement beyond the standard quantum limit (SQL)
\cite{15} and discussed the potential for reaching the Heisenberg limit (HL)
\cite{16}. Zhang et al. \cite{17} proposed an HI with two coherent inputs
and homodyne detection, achieving sub-shot-noise limited phase sensitivity
while retaining robustness against photon loss and background noise. Zhou et
al. \cite{18} showed that the HI can outperform the standard SUI. Chang et
al. \cite{19} demonstrated that the HI not only enhances phase sensitivity
but also effectively resists decoherence induced by photon loss. Despite
these developments, photon loss \cite{20,21} remains a critical obstacle for
practical applications.

To mitigate the detrimental effects of photon loss and further improve phase
sensitivity, non-Gaussian operations have emerged as a powerful strategy.
Techniques including photon catalysis \cite{22,23}, number-conserving
operations \cite{24,25}, photon addition \cite{26,27}, and multi-photon
subtraction (MPS) \cite{28,29,30} have been widely used to improve
interferometric robustness. In particular, MPS operations have been
successfully applied in SUI configurations: Kang et al. \cite{29} enhanced
phase sensitivity via MPS inside the SUI; Li et al. \cite{31} showed that
delocalized photon subtraction can push phase sensitivity closer to the QCRB
and improve loss resistance; Zhou et al. \cite{18} demonstrated that photon
subtraction markedly boosts phase sensitivity, quantum Fisher information
(QFI) \cite{32,33,34}, and noise resilience. However, few studies have
explored MPS operations combined with nonlinear elements within the HI
framework.

In parallel with the development of MPS operations, the integration of
nonlinear phase shifters has emerged as a powerful complementary strategy.
As demonstrated in MZI and SUI configurations, Chang et al. \cite{35} showed
that using a Kerr nonlinear phase shifter in an SUI significantly enhances
phase sensitivity and QFI, even with photon loss. In the lossless case, it
can surpass the conventional Heisenberg scaling $(1/N)$ and approach the
super-Heisenberg scaling $(1/N^{2})$ enabled by the Kerr nonlinearity,
achieving what can be called a super Heisenberg limit (super-HL). Zhao et
al. \cite{27} demonstrated that quantum measurement precision can be
effectively enhanced by employing Kerr nonlinear phase shifters.

The aforementioned works have separately established the merits of the HI
architecture, MPS operations, and Kerr nonlinearities. A natural and
promising extension is to integrate all three elements within a single
framework. To our knowledge, no prior work has combined Kerr nonlinearity,
MPS, and the HI architecture. This integration is promising because the Kerr
effect encodes phase information in higher-order photon moments, MPS tailors
non-Gaussian features to enhance sensitivity, and the HI preserves more
signal photons than an SUI. We hypothesize that their synergy can surpass
existing protocols, including the HI with MPS \cite{18} and the SUI with
Kerr nonlinearity \cite{35}. Physically, the HI replaces the second OPA of
the SUI with a passive BS, thus avoiding the absorption of Kerr-enhanced
probe photons and fully exploiting the nonlinear phase encoding. The MPS
operation then probabilistically sculpts the non-Gaussian photon-number
distribution, concentrating it into those Fock-state components that are
most sensitive to the $\hat{n}^2$ phase shift. Compared with the HI with MPS
(Ref.~\cite{18}) and the SUI with Kerr nonlinearity (Ref.~\cite{35}), our
scheme consistently surpasses the sub-Heisenberg limit and approaches the
super-Heisenberg scaling $\propto 1/N^{2}$, while exhibiting the highest
robustness against photon loss; quantitative comparisons are provided in
Sec.~III.A and Fig.~6. Building upon these previous studies, this work
investigates phase estimation in the HI framework for both ideal and
photon-loss cases. We combine it with MPS operation, utilizing a coherent
state (CS) and a vacuum state (VS) as inputs. In addition, we consider two
specific schemes: Scheme I (linear phase shifter, $k=1$) and Scheme II
(nonlinear Kerr phase shifter, $k=2$). Throughout this paper, we denote the
phase sensitivity for Scheme I and II as $\Delta\phi_1$ and $\Delta\phi_2$,
respectively. We systematically investigate the phase sensitivity, the QFI,
and the QCRB under homodyne detection \cite{36,37}.

This paper is organized as follows: Section II introduces the theoretical
model. Section III presents phase sensitivity analyses under both ideal and
lossy conditions. Section IV investigates the QFI and QCRB in ideal and
lossy scenarios. Section V summarizes our conclusions.

\section{Model}

\begin{figure}[tbh]
\label{Fig1} \centering%
\subfigure{
\begin{minipage}[b]{0.5\textwidth}
\includegraphics[width=0.83\textwidth]{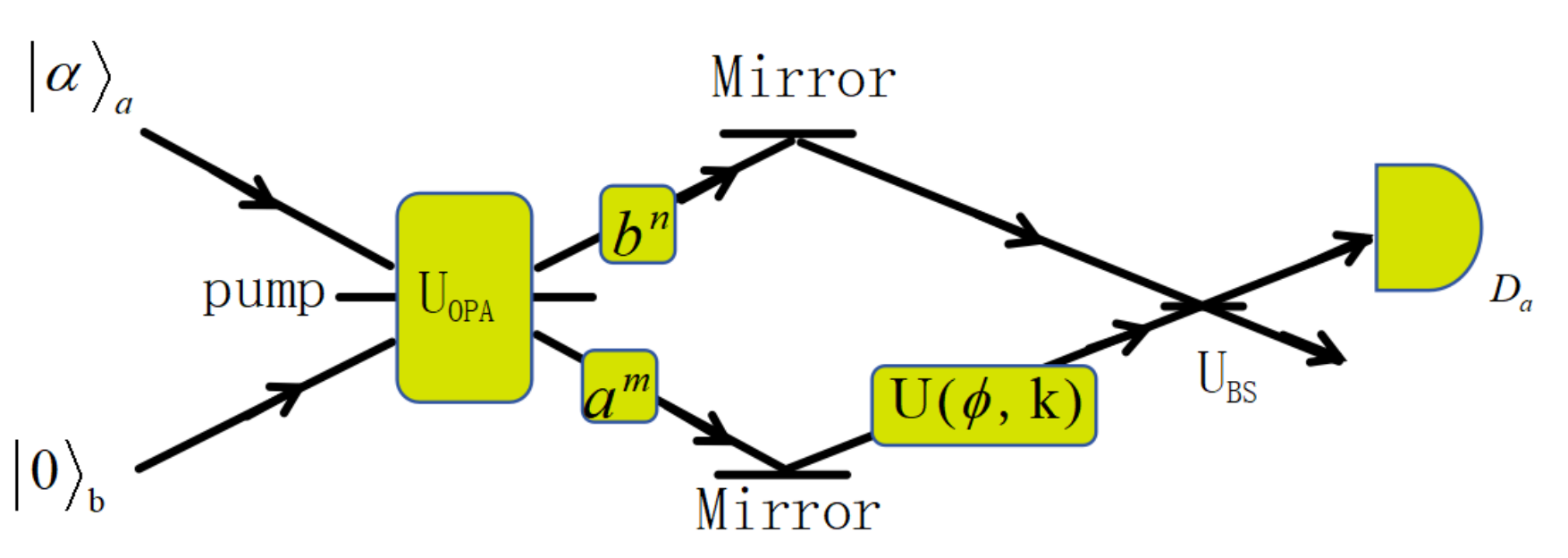}
\end{minipage}}
\caption{Schematic diagram of the proposed hybrid interferometer (HI)
incorporating multi-photon subtraction (MPS). The interferometer is seeded
with a coherent state $\left\vert \protect\alpha \right\rangle _{a}$ and a
vacuum state $\left\vert 0\right\rangle _{b}$. $U_{OPA}$ denotes the optical
parametric amplifier, $U_{BS}$ stands for a 50:50 beam splitter, $U(\protect%
\phi ,k)$ represents the linear or Kerr nonlinear phase shifter, and $D_{a}$
is the homodyne detector on output mode $a$.}
\end{figure}

Figure 1 shows the schematic of the proposed hybrid interferometer (HI)
equipped with a phase shifter and local multi-photon subtraction (MPS). The
HI consists of an OPA, a BS, and a phase shifter. The input states are a
coherent state $\left\vert \alpha \right\rangle _{a}$ (where $\alpha
=\left\vert \alpha \right\vert e^{i\theta _{\alpha }}$ denotes the coherent
amplitude) and a vacuum state $\left\vert 0\right\rangle _{b}$. For
simplicity, we set $\theta _{\alpha }=0$. Next, we briefly comment on the
experimental feasibility of the optical scheme in Fig. 1. All key components
are compatible with current quantum optical technology. For instance, OPA
gains of $g\sim 1-2$ are routinely achieved in periodically poled crystals;
strong Kerr nonlinearities with effective $\chi ^{(3)}$ co-efficients can be
realized in photonic crystal fibers or atomic ensembles; and
photon-number-resolving detectors based on transition-edge sensors now reach
detection efficiencies above $95\%$. Practical imperfections such as phase
diffusion and detector dark counts may degrade performance, but their
influence can be mitigated with post-selection and stabilization techniques.
Specific to our scheme, the Kerr nonlinearity required to achieve the
predicted super-Heisenberg scaling corresponds to an effective $\chi ^{(3)}$
oefficient that induces a phase shift per photon on the order of $10^{-2}$
rad. Such few-photon nonlinearities are becoming accessible in multiple
experimental platforms. For instance, Stolz et al. demonstrated a deterministic 
photon-photon quantum logic gate in a cold atomic ensemble via Rydberg electromagnetically 
induced transparency (EIT), an implementation that fundamentally relies on a quantum-optical 
self-Kerr nonlinearity at the few-photon level \cite{45}. In the microwave domain, Kirchmair et al.,
engineered an artificial Kerr medium using a three-dimensional circuit
quantum electrodynamics architecture and observed the collapse and revival
of a coherent state due to the single-photon Kerr effect \cite{46},
demonstrating that the single-photon Kerr regime---where the interaction
strength between individual photons exceeds the loss rate---is
experimentally attainable. While both Rydberg-EIT and
superconducting-circuit approaches still face challenges in combining high
nonlinearity with low insertion loss for free-space or fiber-based
interferometry, their rapid progress strongly supports the near-term
feasibility of the nonlinear phase shifter assumed in our scheme. The
initial two-mode squeezed vacuum state can be generated using a standard OPA
with a periodically poled nonlinear crystal. The local MPS operation is
implemented via heralded post-selection with low-reflectivity BSs and
photon-number-resolving detectors. The phase shifter can be realized by a
linear electro-optic modulator or, for the Kerr nonlinear case, a nonlinear
medium such as an optical fiber. The interferometer employs a balanced 50:50
BS, and balanced homodyne detection at the output with a phase-locked local
oscillator enables shot-noise-limited quadrature measurements for phase
sensitivity extraction. While all components are within reach, practical
challenges remain, such as achieving high gain $g$ in the OPA,
high-efficiency photon-number-resolving detection for MPS, and sufficiently
strong Kerr nonlinearities; recent experimental progress is steadily
addressing these issues. The proposed scheme relies on well-established
optical components and techniques and is experimentally realizable with
current technology, although the following analysis focuses on the
theoretical performance.

The OPA is described by a two-mode squeezing operator with unitary evolution
$\hat{U}_{OPA}(\xi )=\exp(\xi ^{\ast }\hat{a}\hat{b}-\xi \hat{a}^{\dagger }%
\hat{b}^{\dagger })$, where $\hat{a}(\hat{a}^{\dagger })$ and $\hat{b}(\hat{b%
}^{\dagger }$) denote the photon annihilation (creation) operators for modes
$a$ and $b$, respectively. The squeezing parameter is defined as $\xi
=ge^{i\theta }$, where $g$ is the gain factor and $\theta $ is the phase
parameter. We simplify the analysis by setting $\theta =0$. The phase shift $%
\phi$ in the HI is described by the phase-shifting operator $\hat{U}(\phi
,k)=\exp [i\phi (\hat{a}^{\dagger }\hat{a})^{k}]$, where $k$ is a
real-valued parameter. In this study, we consider two schemes: the linear
phase shifter $(k=1)$ denoted as Scheme I, and the nonlinear Kerr phase
shifter $(k=2)$ denoted as Scheme II. The corresponding unitary operator for
the BS is $\hat{U}_{BS}=\exp (\frac{-i\pi }{4}( \hat{a}^{\dagger }\hat{b}+%
\hat{a}\hat{b}^{\dagger }) )$. The operation $\hat{a}^{m}\hat{b}^{n}$
represents local MPS, where $m$ photons are subtracted from mode $a$ and $n$
photons are subtracted from mode $b$ independently. The two subtraction
operations act locally on each mode and are path-distinguishable. Phase
estimation is performed via homodyne detection on output mode $a$ at the
output. The final output state after the interferometer is thus given by:
\begin{equation}
\left\vert \psi _{out}\right\rangle =\lambda \hat{U}_{BS}\hat{U}(\phi ,k)%
\hat{a}^{m}\hat{b}^{n}\hat{U}_{OPA}\left\vert \psi _{in}\right\rangle ,
\end{equation}%
where the input state is defined as $\left\vert \psi _{in}\right\rangle
=\left\vert \alpha \right\rangle _{a}\left\vert 0\right\rangle _{b}$. The
corresponding transformations for Scheme I and Scheme II are given by:

\begin{equation}
\hat{U}^{\dagger }(\phi ,1)\hat{a}\hat{U}(\phi ,1)=\hat{a}e^{i\phi },
\label{eq:linear}
\end{equation}

\begin{equation}
\hat{U}^{\dagger }(\phi ,2)\hat{a}\hat{U}(\phi ,2)=e^{i\phi }e^{2i\phi \hat{%
a }^{\dagger }\hat{a}}\hat{a},  \label{eq:kerr}
\end{equation}%
To compute expectation values of observables (e.g., $\left\langle \hat{X}%
_{a}\right\rangle $), which involve sequences of creation and annihilation
operators, we employ a generating function method. This allows systematic
calculation of quantities like $\left\langle \psi _{out}\right\vert \hat{a}%
^{\dagger x1}\hat{a}^{y1}\hat{b}^{\dagger x2}\hat{b}^{y2}\left\vert \psi
_{out}\right\rangle $ by differentiating a generating function $%
Q_{mn,x1,y1,x2,y2}.$ The normalization constant $\lambda $ and the
generating function $Q$ are defined as:
\begin{equation}
\left\vert \lambda \right\vert ^{2}=\frac{1}{Q_{mn,0,0,0,0}},
\end{equation}%
where
\begin{eqnarray}
Q_{mn,x1,y1,x2,y2} &=&\frac{\partial ^{2m+2n+x1+y1+x2+y2}}{\partial \lambda
_{1}^{m}\partial \lambda _{2}^{n}\partial \lambda _{3}^{x1}\partial \lambda
_{4}^{y1}\partial \lambda _{5}^{x2}\partial \lambda _{6}^{y2}\partial
\lambda _{7}^{m}\partial \lambda _{8}^{n}}  \notag \\
&&e^{w_{4}}|_{\lambda _{1}=\lambda _{2}=\lambda _{3}=\lambda _{4}=\lambda
_{5}=\lambda _{6}=\lambda _{7}=\lambda _{8}=0},
\end{eqnarray}%
with

\begin{eqnarray}
w1 &=&\left[ \left( \lambda _{2}+\lambda _{7}\right) \cosh g-\left( \lambda
_{3}+\lambda _{6}\right) \sinh g\right] ,  \notag \\
w2 &=&\left[ \left( \lambda _{1}+\lambda _{5}\right) \cosh g-\left( \lambda
_{4}+\lambda _{8}\right) \sinh g\right] ,  \notag \\
w3 &=&\left( \lambda _{1}+\lambda _{5}\right) \sinh g\times  \notag \\
&&\left[ \left( \lambda _{2}+\lambda _{7}\right) \sinh g-\left( \lambda
_{3}+\lambda _{6}\right) \cosh g\right]  \notag \\
&&-\left( \lambda _{4}+\lambda _{8}\right) \sinh g\times w1,  \notag \\
w4 &=&w1\alpha +w2\alpha ^{\ast }+w3.
\end{eqnarray}%
Here, $x1,y1,x2$ and $y2$ are positive integers, and $\lambda _{i}$ ($%
i=1,2,3,4,5,6,7,8$) are auxiliary differential variables. After
differentiation, all $\lambda _{i}$ are set to zero. The expectation value
of any operator product is then obtained as $\lambda ^{2}Q_{mn,x1,y1,x2,y2}.$

\section{Phase Sensitivity}

In the HI configuration, homodyne detection typically yields superior phase
sensitivity compared with intensity detection. It is experimentally
accessible and theoretically tractable, making it well suited for
continuous-variable quantum metrology and quantum key distribution \cite%
{38,39,40}. We therefore employ homodyne detection on output mode $a$,
measuring the quadrature operator $\hat{X}_{a}=(\hat{a}+\hat{a}^{\dag })/%
\sqrt{2}$, to estimate the unknown phase $\phi $. The phase sensitivity $%
\Delta \phi $ is given by the error-propagation equation \cite{41,42,43}:

\begin{equation}
\Delta \phi =\frac{\sqrt{\left\langle \Delta ^{2}\hat{X}_{a}\right\rangle }}{%
\left\vert \partial \left\langle \hat{X}_{a}\right\rangle /\partial \phi
\right\vert }.
\end{equation}%
Utilizing Eqs.~(1), (2), (3), and (6) together with the generating function
method, the phase sensitivity $\Delta \phi $ of the HI configuration can be
derived analytically. Detailed expressions for $\left\langle \hat{X}%
_{a}\right\rangle $ and $\left\langle \Delta ^{2}\hat{X}_{a}\right\rangle $
for both schemes are provided in Appendix A.

\subsection{Phase sensitivity under Ideal Conditions}

\begin{figure}[tbh]
\label{Fig2} \centering%
\subfigure{
\begin{minipage}[b]{0.5\textwidth}
\includegraphics[width=0.83\textwidth]{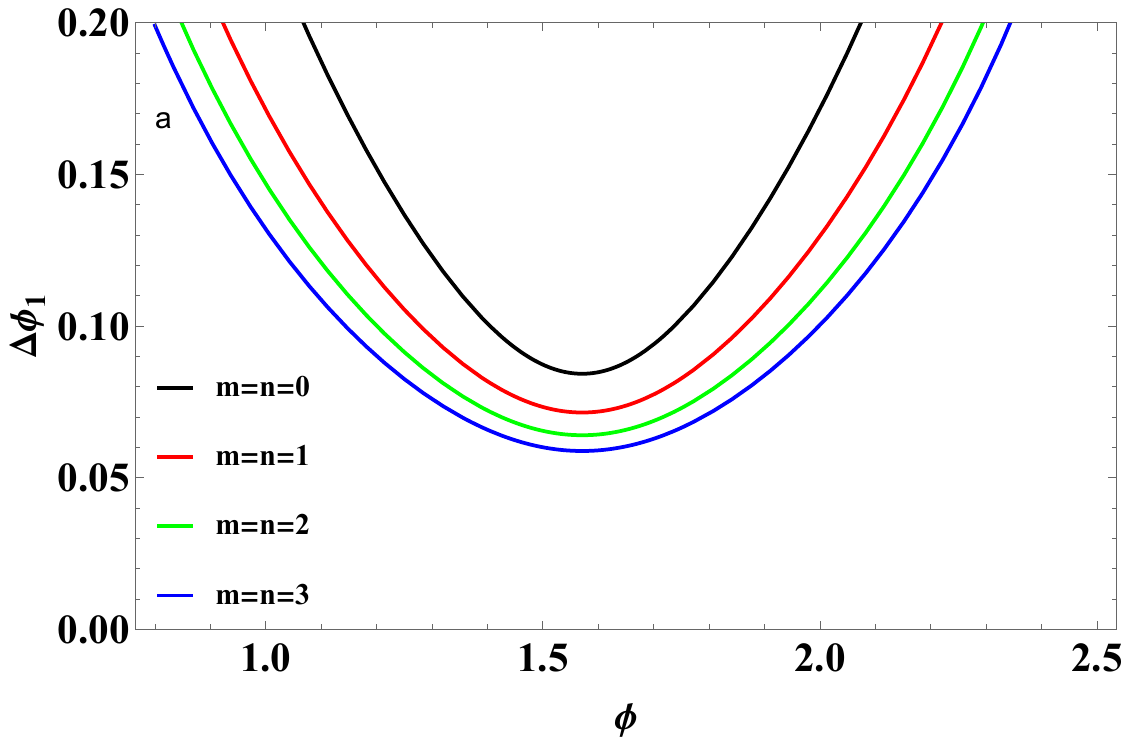}\\
\includegraphics[width=0.83\textwidth]{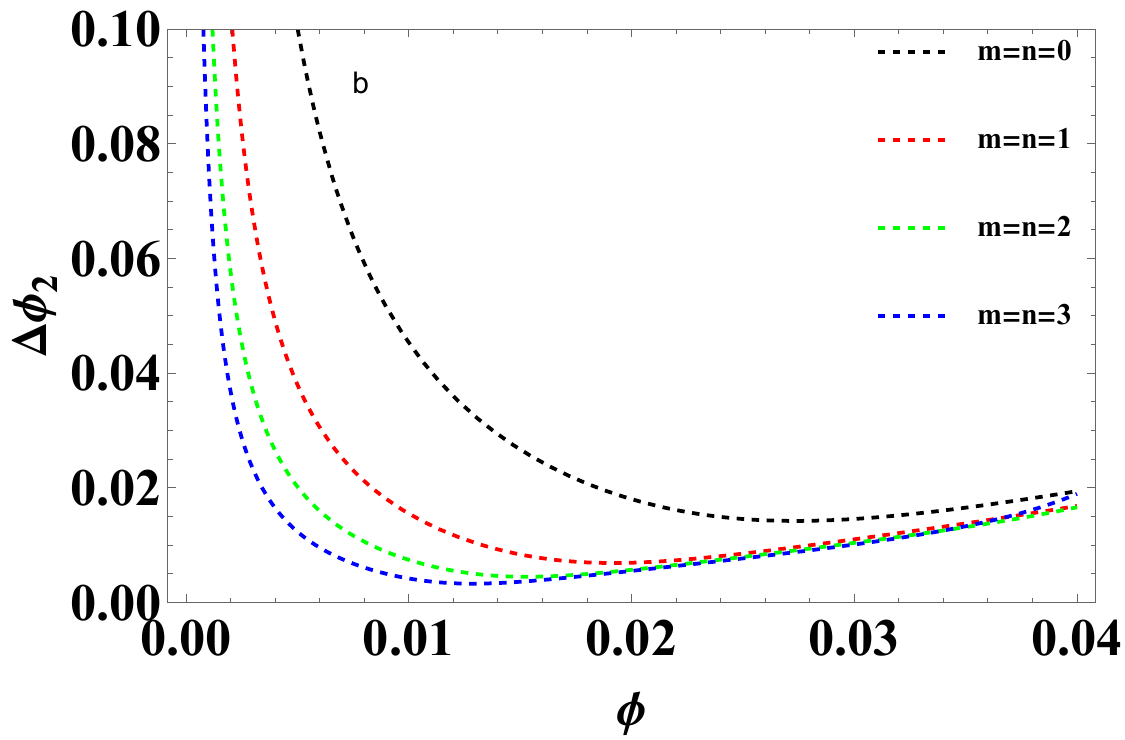}
\end{minipage}}
\caption{Phase sensitivity $\Delta \protect\phi _{1}$ (a) and $\Delta
\protect\phi _{2}$ (b)\ of the HI obtained via homodyne detection as a
function of the phase shift $\protect\phi $ for photon-subtraction orders $%
m=n=0$ (black), $1$ (red), $2$ (green), and $3$ (blue). The parameters are
set as $\protect\alpha =2$ and $g=1$. (a) Linear phase shifter (Scheme I, $%
k=1$). (b) Kerr nonlinear phase shifter (Scheme II, $k=2$).}
\end{figure}

We first consider the ideal lossless scenario $\left( \eta =1\right) $.
Fig.~2 presents the phase sensitivity $\Delta \phi _{1}$ (a) and $\Delta
\phi _{2}$ (b)\ of the HI as a function of phase shift $\phi $ for varying
photon-subtraction orders\ $m=n=0,1,2,3,$\ with the coherent amplitude and
OPA gain factor fixed at $\alpha =2$ and $g=1$. Fig.~2(a) displays results
for the linear phase shifter (Scheme I, $k=1$), while Fig.~2(b) corresponds
to the Kerr nonlinear phase shifter (Scheme II, $k=2$). For both schemes,
the phase sensitivity first improves (decreases) and then deteriorates as $%
\phi $ increases. As expected, smaller values of $\Delta \phi $ correspond
to higher estimation precision. Increasing the number of subtracted photons
shifts the optimal point to a specific phase value and significantly
enhances the overall precision. Notably, Scheme II achieves a much sharper
minimum and far superior phase sensitivity compared to Scheme I. The optimal
(minimum) phase sensitivity is approximately $2.5\times 10^{-3}$ for Scheme
II, in stark contrast to approximately $0.06$ for Scheme I. For Scheme II,
the optimal phase shift is $\phi $ $\approx 0.013$ rad, a value much smaller
than the typical $\sim 1.6$ rad for Scheme I. This reflects the fact that
the Kerr nonlinearity $\propto $ $\hat{n}^{2}$ produces a phase sensitivity
that grows rapidly with photon number, so the homodyne signal changes most
sharply at very small phases when combined with the photon-number
concentration from MPS. This dramatic improvement arises from the
fundamental nature of the Kerr nonlinear phase shifter. The Kerr effect
introduces an intensity-dependent phase shift proportional to $(\hat{a}%
^{\dagger }\hat{a})^{2}.$ This nonlinear response renders the homodyne
signal $\left\langle \hat{X}_{a}\right\rangle $ far more sensitive to
changes in $\phi $ (i.e., yielding a larger $\left\vert \partial
\left\langle \hat{X}_{a}\right\rangle /\partial \phi \right\vert $) around
the optimal point compared to the linear phase shifter, while the noise $%
\left\langle \Delta ^{2}\hat{X}_{a}\right\rangle $ remains comparatively
suppressed. According to Eq. (6), this combination of a larger signal
derivative and controlled noise leads to a significantly lower $\Delta \phi $%
. The MPS operation further enhances this process by non-Gaussianizing the
quantum state, effectively tailoring its photon number distribution to
maximize the benefit of the nonlinear phase sensing.

\begin{figure}[tbh]
\label{Fig3} \centering%
\subfigure{
\begin{minipage}[b]{0.5\textwidth}
\includegraphics[width=0.83\textwidth]{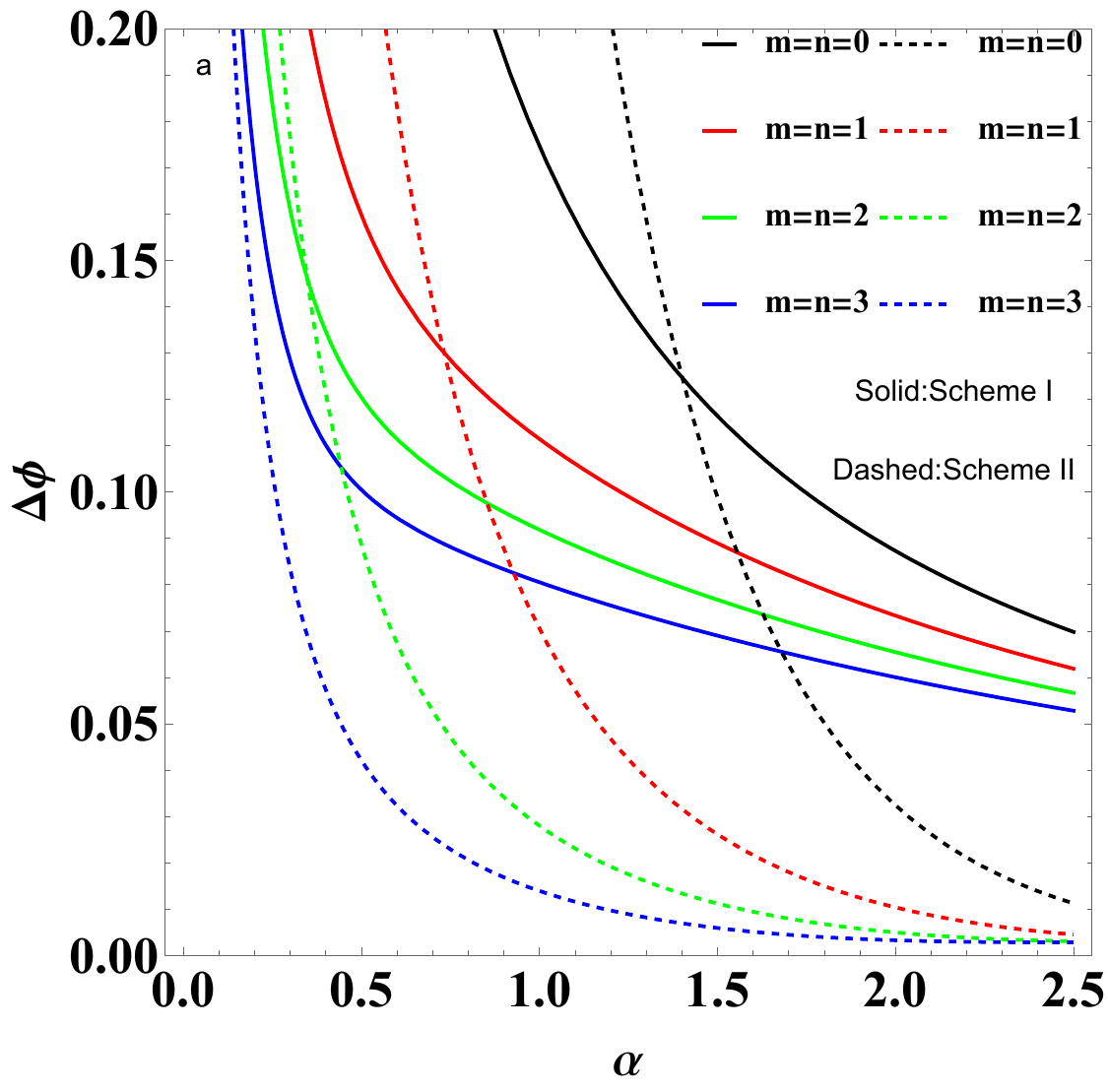}\\
\includegraphics[width=0.83\textwidth]{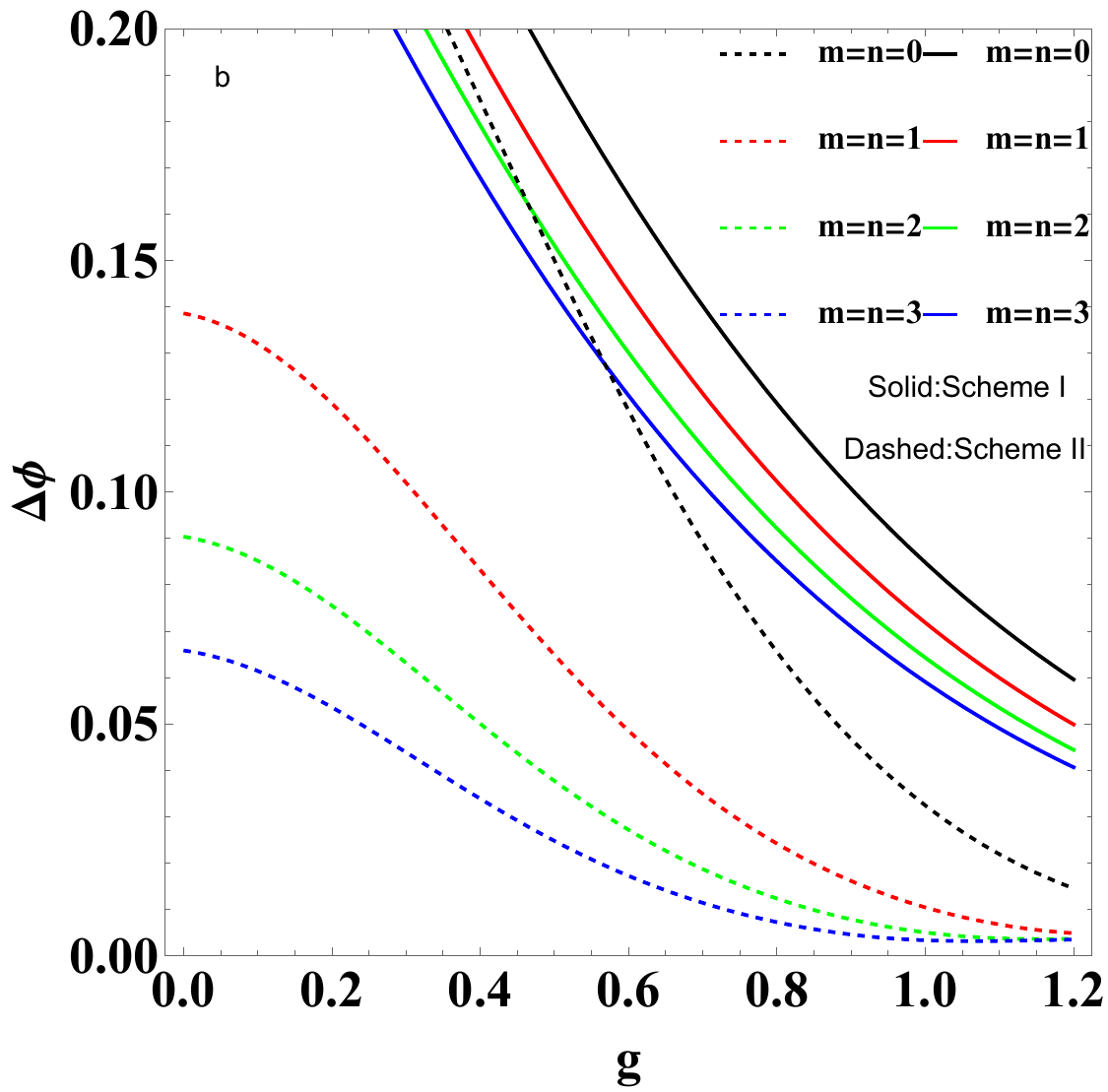}
\end{minipage}}
\caption{Optimal phase sensitivity of the HI for Scheme I (solid lines) and
Scheme II (dashed lines) as functions of (a) the coherent amplitude $\protect%
\alpha $ with $g=1$, and (b) the OPA gain factor $g$ with $\protect\alpha =2$%
. The black, red, green, and blue lines correspond to photon-subtraction
orders $m=n=0$, $1$, $2$ and $3$, respectively.}
\end{figure}

Next, we investigate the dependence of the optimal phase sensitivity on the
key experimental parameters: the coherent amplitude $\alpha $ and the OPA
gain factor $g$. For each set of parameters, the optimal phase $\phi $ that
minimizes $\Delta \phi $ is determined by numerical scanning. Fig.~3
presents the minimum achievable phase sensitivity for Scheme I (solid lines)
and Scheme II (dashed lines) under ideal conditions. Curves corresponding to
different photon-subtraction orders $m=n=0,1,2,3$ are plotted in black, red,
green, and blue, respectively. As shown in Fig.~3(a), the phase sensitivity
improves monotonically with increasing $\alpha $ for both schemes, as a
larger coherent amplitude injects more photons into the interferometer,
thereby increasing the available resources to carry phase information.
Similarly, Fig.~3(b) shows that phase sensitivity also improves with
increasing $g$, a behavior that can be attributed to the signal
amplification provided by the squeezed light generated in the OPA. For
Scheme II, however, this improvement gradually saturates for $g>1$, which
justifies our choice of $g=1$ as a representative operating point for
subsequent detailed analyses. In both cases, the MPS operation further
amplifies this improvement, with the enhancement becoming more pronounced as
the subtraction order increases. Across the entire parameter space, Scheme
II consistently and significantly outperforms Scheme I, unequivocally
confirming the pronounced metrological advantage conferred by the Kerr
nonlinear phase shifter.

To contextualize the performance of the nonlinear Kerr phase shifter (Scheme
II, $k=2)$, we compare its phase sensitivity against several fundamental
quantum limits: the standard quantum limit (SQL, $\Delta \phi _{SQL}=1/\sqrt{%
N}$), the Heisenberg limit (HL, $\Delta \phi _{HL}=1/N$), the sub-Heisenberg
limit (sub-HL, $\Delta \phi _{sub-HL}=1/N^{3/2}$), and the super-Heisenberg
limit (SHL, $\Delta \phi _{SHL}=1/N^{2}$). The SQL and the conventional HL
are the benchmarks for linear ($k=1$) phase accumulation, while the sub-HL
and the SHL are the ultimate scalings allowed by the $k=2$ Kerr
nonlinearity. Here, $N$ denotes the total mean photon number inside the
interferometer before the BS, which is given by:

\begin{eqnarray}
N &=&\lambda ^{2}\left\langle \psi _{in}\right\vert \hat{U}_{OPA}^{\dagger }%
\hat{a}^{\dagger m}\hat{b}^{\dagger n}\hat{U}^{\dagger }(\phi ,k)\left( \hat{%
a}^{\dagger }\hat{a}+\hat{b}^{\dagger }\hat{b}\right)  \notag \\
&&\hat{U}(\phi ,k)\hat{a}^{m}\hat{b}^{n}\hat{U}_{OPA}\left\vert \psi
_{in}\right\rangle  \notag \\
&=&\lambda ^{2}\left( Q_{mn,1,1,0,0}+Q_{mn,0,0,1,1}\right) .
\end{eqnarray}

\begin{figure}[tbh]
\label{Fig4} \centering%
\subfigure{
\begin{minipage}[b]{0.5\textwidth}
\includegraphics[width=0.83\textwidth]{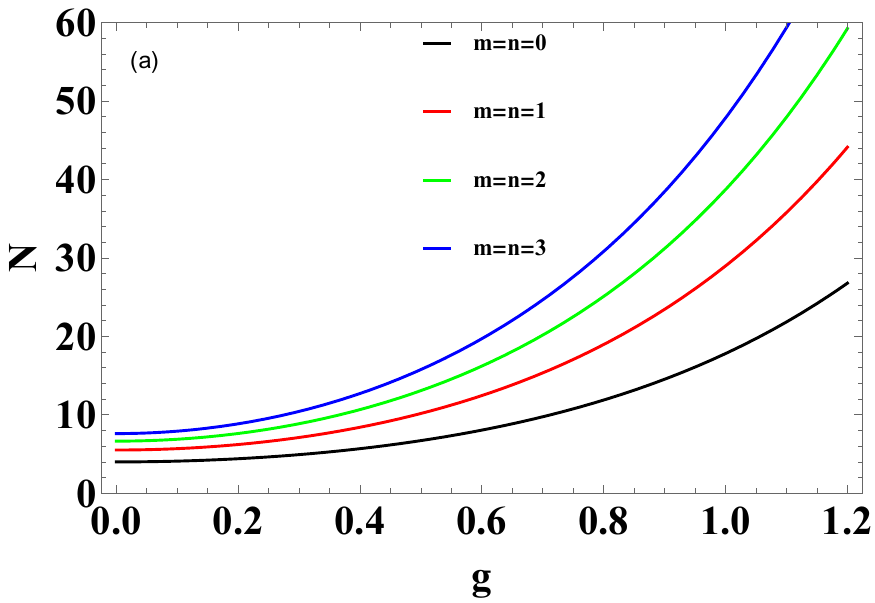}\\
\includegraphics[width=0.83\textwidth]{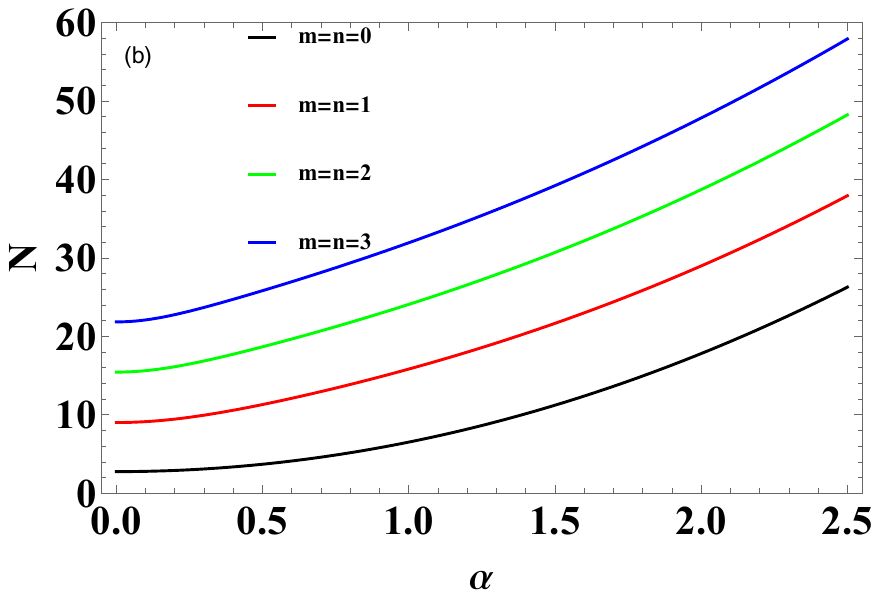}
\end{minipage}}
\caption{The total mean photon number $N$ in the HI configuration as a
function of (a) the gain factor $g$ with $\protect\alpha =2$, and (b) the
coherent amplitude $\protect\alpha $ with $g=1$. The black, red, green, and
blue lines correspond to photon-subtraction orders $m=n=0,1,2$ and $3$,
respectively.}
\end{figure}

\begin{figure}[tbh]
\label{Fig5} \centering%
\subfigure{
\begin{minipage}[b]{0.5\textwidth}
\includegraphics[width=0.83\textwidth]{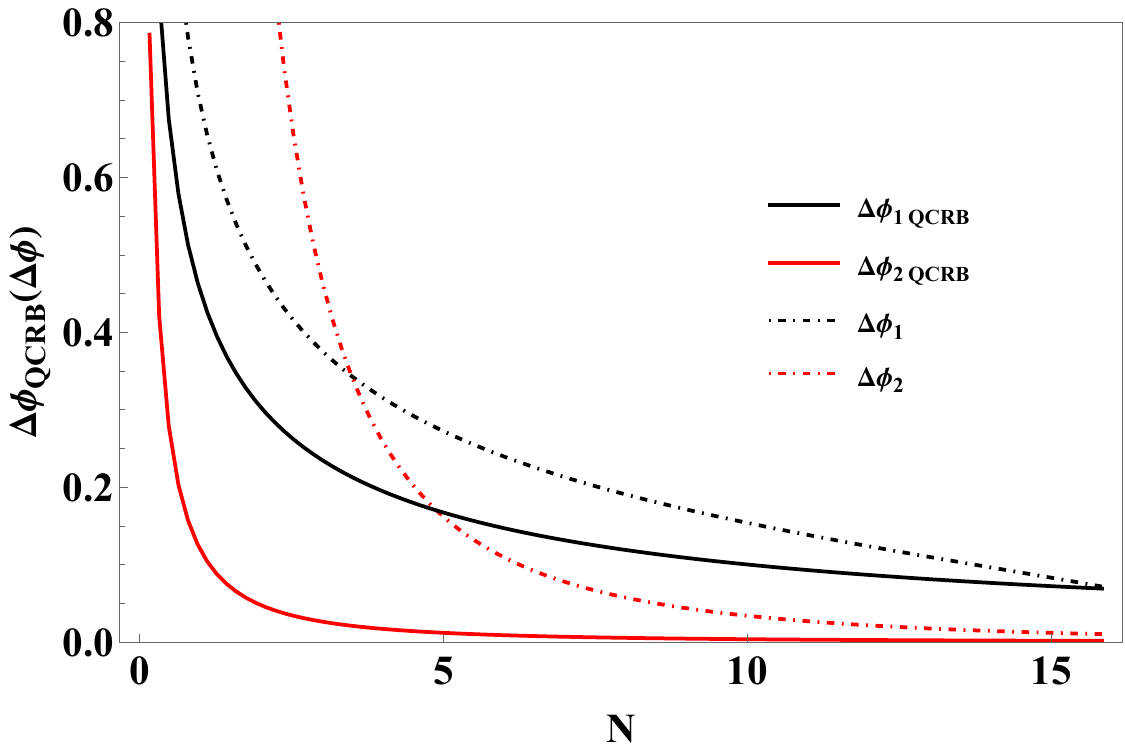}
\end{minipage}}
\caption{Phase sensitivity (dashed) and QCRB (solid) as functions of $N$ for
Scheme I ($\protect\phi =1.6$, black) and Scheme II ($\protect\phi =0.013$,
red). The QFI and QCRB are independent of $\protect\phi $ and are evaluated
on the pre-BS state $|\Psi _{\protect\phi }\rangle $; The optimal $\protect%
\phi $ values indicated above are those that minimize the homodyne phase
sensitivity and are used only for labeling the dashed curves. The optimal $%
\protect\phi $ values are chosen to minimize $\Delta \protect\phi $ for each
scheme. Parameters are fixed at $\protect\alpha =2$ and $g=1$.}
\end{figure}

To further elucidate how the OPA gain factor $g$ and the coherent amplitude $%
\alpha $ modulate the total mean internal photon number $N$ of the HI, we
plot $N$ as a function of $g$ and $\alpha $ in Fig.~4. As physically
expected, $N$ rises markedly with both parameters, and this growth is
further enhanced as the photon-subtraction order increases. Although a
larger mean photon number generally offers more quantum resources for phase
encoding and estimation, the correlation between $N$ and phase sensitivity
remains nontrivial and requires systematic analysis. We therefore
characterize this inherent relationship in Fig.~5, which plots the phase
sensitivity and its QCRB as functions of $N$ for Scheme I and Scheme II. The
results confirm that the phase sensitivity improves significantly (i.e., $%
\Delta \phi $ decreases) with increasing $N$, consistent with the intuition
that a larger photon resource budget enhances estimation precision. However,
this improvement gradually saturates at high $N$, where the curves flatten
and show negligible further reduction. This saturation arises from two key
mechanisms: the practical sensitivity approaches the fundamental quantum
limit set by the QCRB, where the QFI scaling with $N$ begins to saturate due
to the finite photon resources available in the probe state; simultaneously,
the signal-to-noise ratio improvement plateaus, as the phase-dependent slope
of the homodyne signal grows more slowly than $N$ itself, and even the Kerr
nonlinear enhancement cannot sustain the initial rapid precision gains
indefinitely. Across all $N$, Scheme II consistently outperforms Scheme I,
demonstrating the advantage of nonlinear phase sensing for enhanced
sensitivity. The close agreement between the dashed and solid lines for both
schemes indicates that homodyne detection achieves a phase sensitivity near
the fundamental QCRB, i.e., the measurement is nearly optimal.
\begin{figure}[tbh]
\label{Fig6} \centering%
\subfigure{
\begin{minipage}[b]{0.5\textwidth}
\includegraphics[width=0.83\textwidth]{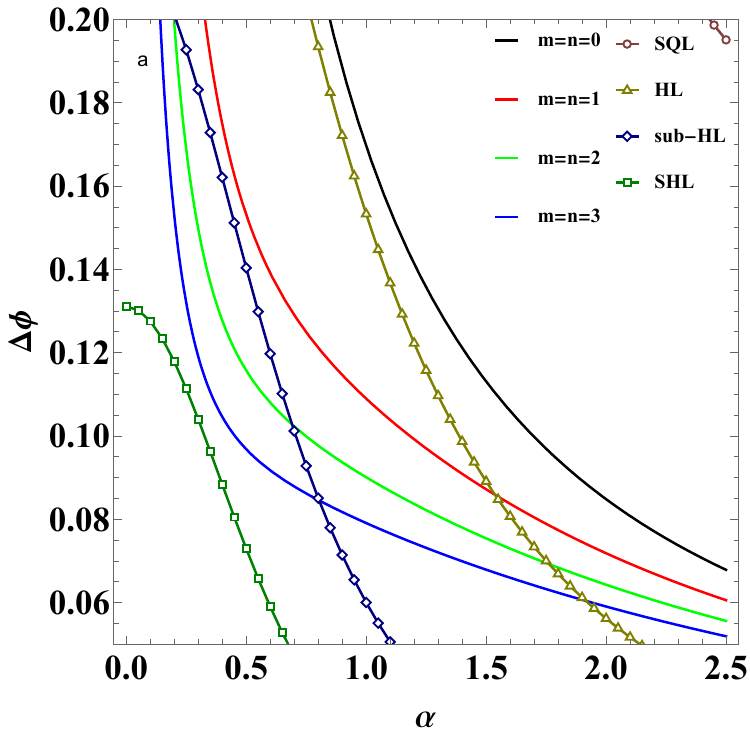}\\
\includegraphics[width=0.83\textwidth]{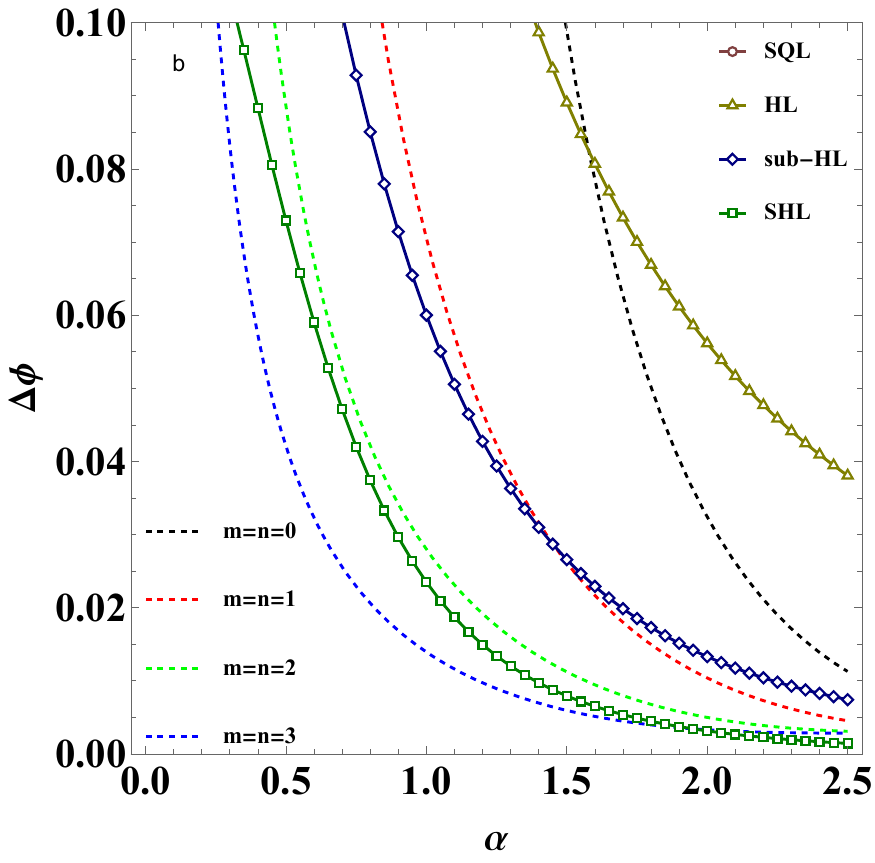}
\end{minipage}}
\caption{Phase sensitivity of the HI with MPS and fundamental quantum limits
versus the coherent amplitude $\protect\alpha $ with $g=1$. (a) Linear phase
shifter (Scheme I). (b) Kerr nonlinear phase shifter (Scheme II). Circles
mark the SQL, triangles mark the conventional HL, diamonds mark the sub-HL,
and squares mark the SHL.}
\end{figure}

We stress that the super-Heisenberg scaling $1/N^{2}$ observed here arises
from the nonlinear Kerr Hamiltonian $\propto $ $(\hat{a}^{\dagger }\hat{a}%
)^{2}$ and does not contradict the fundamental Heisenberg limit for linear
phase accumulation. For linear phase shifts, the ultimate scaling is $1/N$;
the Kerr-enhanced scaling is a result of the specific nonlinear dynamics and
the associated quantum Fisher information scaling.

As illustrated in Fig. 6, we systematically analyze the influence of $\alpha
$ on the phase sensitivity $\Delta \phi $ and bench mark its performance
against fundamental measurement precision limits under ideal conditions.
Fig.~6(a) presents the phase sensitivity of the linear phase shifter (Scheme
I). As $\alpha $ increases, $\Delta \phi $ decreases rapidly, indicating a
steady improvement in measurement precision. Notably, regardless of which
MPS operation is employed, Scheme I consistently outperforms the standard
quantum limit (SQL) but fails to reach the super-Heisenberg limit (SHL). In
the absence of MPS, Scheme I cannot surpass the conventional HL; However,
with the implementation of MPS, the phase sensitivity exceeds the HL over an
extended range of $\alpha $, and this range expands as the number of
subtracted photons increases. The sub-HL is only surpassed when at least two
photons are subtracted, with the applicable $\alpha $ range further widening
as the subtraction order increases. In stark contrast, Fig.~6(b) depicts the
phase sensitivity of Scheme II, which exhibits substantially superior
performance. Even without MPS, Scheme II surpasses the conventional HL
within a specific range of $\alpha $. With the integration of MPS, it
consistently exceeds the sub-HL over an extended $\alpha $ range, which
broadens as the number of subtracted photons increases. Only when three
photons are subtracted does the phase sensitivity approach the SHL. This
remarkable enhancement stems from the Kerr nonlinear phase shifter, which,
when combined with MPS, more effectively extracts and amplifies information
about the unknown phase $\phi $ compared to the linear phase shifter. To
further highlight the advantage of our proposed combined scheme, we compare
it with two key recent studies in the field. Compared to the HI with MPS
(similar to our Scheme I) reported in Ref.~\cite{18}, our Scheme II achieves
a lower phase sensitivity $\Delta \phi $ for the same mean photon number $N$
and subtraction order, particularly in the regime approaching the SHL.
Furthermore, in comparison to the SUI with a Kerr nonlinear phase shifter in
Ref.~\cite{35}, our scheme within the HI architecture demonstrates superior
capability in surpassing the HL and sub-HL. Physically, in an SUI the second
OPA inevitably absorbs some of the probe photons that encode the
Kerr-modified phase, thereby diluting the nonlinear advantage. The HI
replaces the second OPA with a passive BS, preserving these photons and
allowing the full benefit of the Kerr nonlinearity to be exploited at the
detector. Importantly, it also leverages the inherent photon-recycling
advantage of the HI over the SUI. This comparative analysis underscores the
synergistic benefit of integrating Kerr nonlinearity and MPS within the
optimized HI framework, validating the superiority of our proposed scheme.

\subsection{Phase sensitivity under Photon Loss}

\begin{figure}[tbh]
\label{Fig7} \centering%
\subfigure{
\begin{minipage}[b]{0.5\textwidth}
\includegraphics[width=0.83\textwidth]{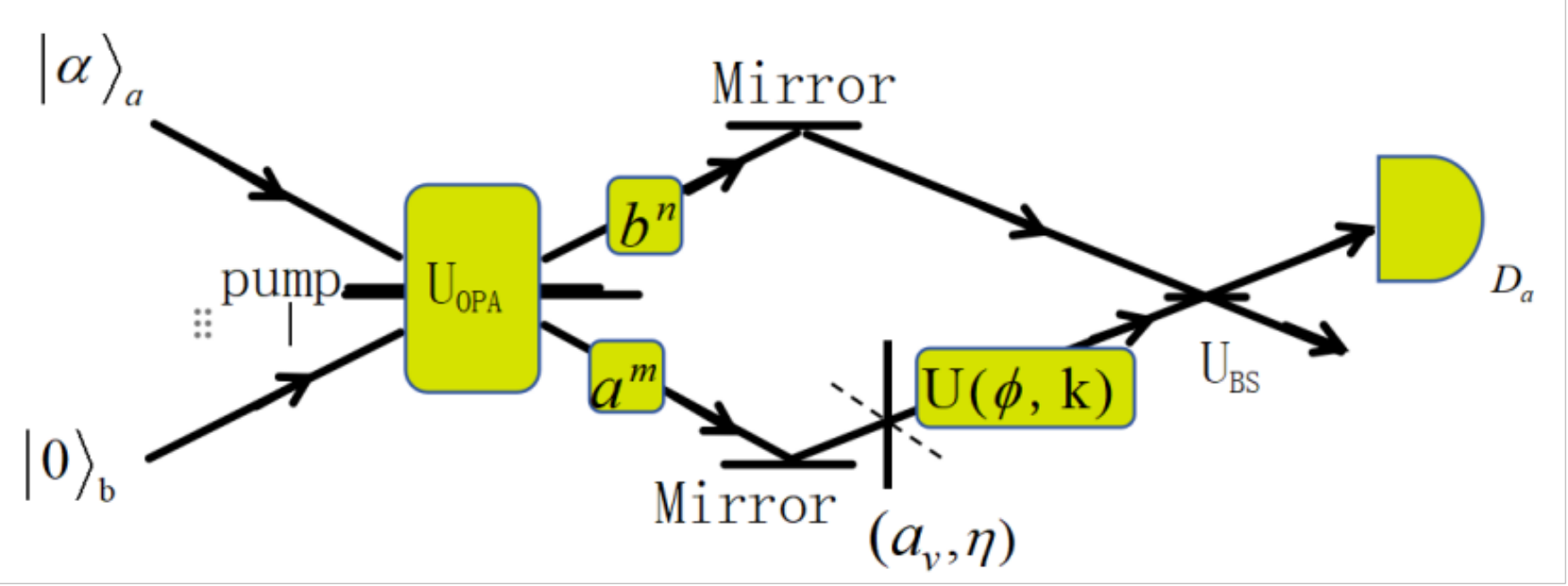}
\end{minipage}}
\caption{Schematic of single-mode internal photon loss in mode $a$, modeled
by a fictitious beam splitter with transmissivity $\protect\eta $ positioned
before the phase shifter. The auxiliary vacuum mode $\hat{a}_{v}$ accounts
for photons lost to the environment.}
\end{figure}

In real-world experimental scenarios, systems inevitably interact with their
surrounding environments, leading to photon loss, which degrades fidelity
and causes phase diffusion. Such internal losses are particularly
detrimental and unavoidable. In this section, we investigate the impact of
internal photon losses on the phase sensitivity for Schemes I and II. To
model internal photon loss, we consider losses occurring in arm $a$ before
the phase shifter, which is implemented using a fictitious beam splitter
with transmissivity $\eta $, as depicted in Fig.~7. The unitary operator
describing this loss process is given by $\hat{U}_{La}=\exp [(\hat{a}%
^{\dagger }\hat{a}_{v}-\hat{a}\hat{a}_{v}^{\dagger })\arccos \sqrt{\eta }]$,
where $\hat{a}_{v}$ denotes an auxiliary vacuum mode that accounts for
photons lost to the environment. As expected, the ideal lossless scenario
corresponds to $\eta $ $=1$, where no photons are scattered or absorbed by
the environment.

\begin{figure}[tbh]
\label{Fig8} \centering%
\subfigure{
\begin{minipage}[b]{0.5\textwidth}
\includegraphics[width=0.83\textwidth]{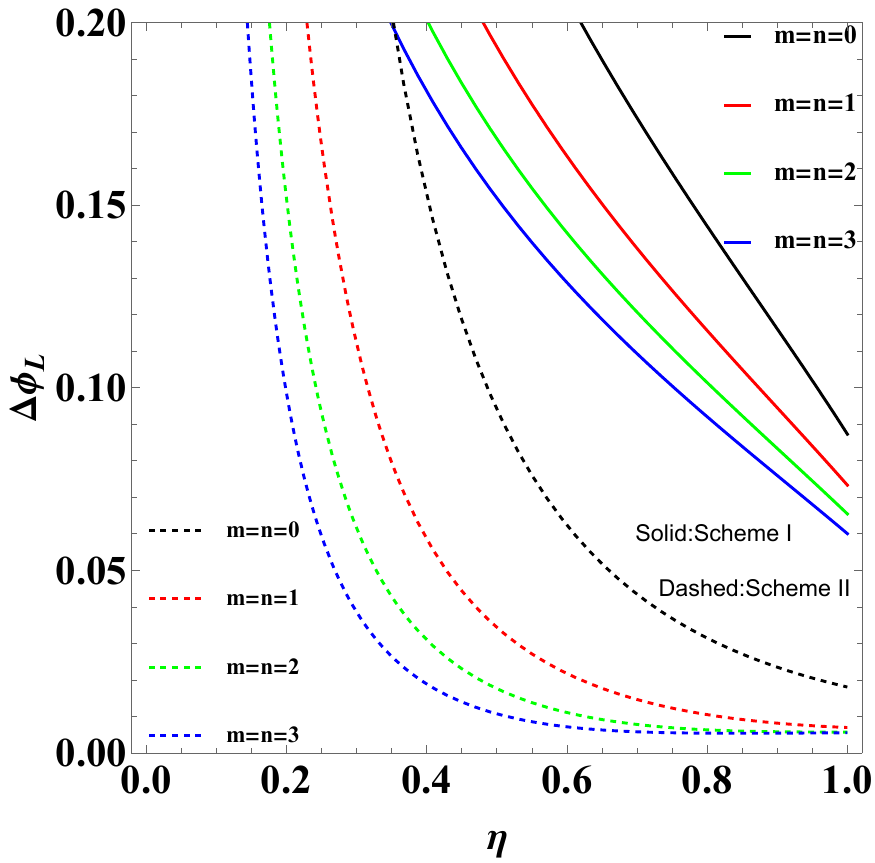}
\end{minipage}}
\caption{Optimal phase sensitivity of the HI as a function of transmissivity
$\protect\eta $ with $\protect\alpha =2$, $g=1$, $\protect\phi =1.6$ for
Scheme I, and $\protect\phi =0.013$ for Scheme II. Note that decreasing $%
\protect\eta $ corresponds to increased photon loss.}
\end{figure}

Fig.~8 presents the optimal phase sensitivity $\Delta \phi _{1}$ (Scheme I)
and $\Delta \phi _{2}$ (Scheme II) as functions of the transmissivity $\eta $
for MPS operations. The simulations are performed with fixed parameters: $%
\alpha =2$, $g=1$, $\phi =1.6$ for Scheme I, and $\phi =0.013$ for Scheme
II. As $\eta $ decreases, the degree of internal photon loss increases,
which is consistent with the physical definition of transmissivity, where
lower $\eta $ corresponds to higher photon attenuation. As physically
expected, the optimal phase sensitivity improves (i.e., $\Delta \phi $
decreases) with increasing $\eta $ for both schemes, and this improvement
becomes more pronounced as the number of subtracted photons increases.
Importantly, Scheme II consistently and significantly outperforms Scheme I
across the entire range of $\eta $, demonstrating superior robustness
against internal photon loss. This enhanced loss resilience stems from the
synergistic action of the Kerr nonlinearity and MPS operation. Specifically,
the MPS operation engineers a non-Gaussian quantum state, whose phase
sensitivity is inherently less vulnerable to the decohering effects of
photon loss. Meanwhile, the Kerr nonlinear phase shifter induces a phase
shift proportional to $\hat{n}_{a}^{2}$ (where $\hat{n}_{a}=\hat{a}^{\dagger
}\hat{a}$ is the photon number operator for mode $a$). Even when photon loss
reduces the average photon number in mode $a$, the nonlinear phase signal
retains a stronger dependence on the fluctuations of the remaining photon
number compared to the linear phase signal (proportional to $\hat{n}_{a}$ in
Scheme I). This intrinsic property renders the measurement more resilient to
attenuation. Together, the synergy between these two effects suppresses
decoherence far more effectively than either the Kerr nonlinearity or MPS
alone could achieve.

\section{The Quantum Cram\'{e}r-Rao Bound}

The preceding section analyzed the phase sensitivity $\Delta \phi $
achievable with homodyne detection. We now shift our focus to the QFI, which
quantifies the maximum information about the phase $\phi $ that can be
extracted from the quantum state itself, independent of the measurement
scheme. The QFI directly dictates the ultimate precision bound for phase
estimation, known as the quantum Cram\'{e}r-Rao bound (QCRB), which is given
by:

\begin{equation}
\Delta \phi _{QCRB}=\frac{1}{\sqrt{F_{k}}} \quad (k=1,2).
\end{equation}%
where $k=1$ and $k=2$ correspond to Scheme I and Scheme II, respectively.

\subsection{The QCRB under Ideal Conditions}

For a pure state $\left\vert \Psi _{\phi }\right\rangle $ in the absence of
loss, the QFI is given by%
\begin{equation}
F_{k}=4\left[ \left\langle \Psi _{\phi }^{\prime }\right\vert \Psi _{\phi
}^{\prime }\rangle -\left\vert \left\langle \Psi _{\phi }^{\prime
}\right\vert \Psi _{\phi }\rangle \right\vert ^{2}\right] ,
\end{equation}%
where $\left\vert \Psi _{\phi }\right\rangle =\lambda \hat{U}(\phi ,k)\hat{a}%
^{m}\hat{b}^{n}\hat{U}_{OPA}\left\vert \psi _{in}\right\rangle $ denotes the
state after the phase shifter and before the beam splitter. Here, $%
\left\vert \Psi _{\phi }^{\prime }\right\rangle =\partial \left\vert \Psi
_{\phi }\right\rangle /\partial \phi $. For the linear phase shifter ($k=1)$%
, the QFI simplifies to the variance of the photon number in mode $a$:

\begin{equation}
F_{1}=4\langle \Delta ^{2}\hat{n}_{a}\rangle .
\end{equation}%
where
\begin{equation}
\langle \Delta ^{2}\hat{n}_{a}\rangle =\left\langle \Psi _{\phi }\right\vert
\left( \hat{a}^{\dagger }\hat{a}\right) ^{2}\left\vert \Psi _{\phi
}\right\rangle -\left( \left\langle \Psi _{\phi }\right\vert \hat{a}%
^{\dagger }\hat{a}\left\vert \Psi _{\phi }\right\rangle \right) ^{2}.
\end{equation}%
Using the generating function, this becomes
\begin{eqnarray}
F_{1} &=&4[\lambda ^{2}Q_{mn,2,2,0,0}+\lambda ^{2}Q_{mn,1,1,0,0}-  \notag \\
&&(\lambda ^{2}Q_{mn,1,1,0,0})^{2}],
\end{eqnarray}%
For the Kerr nonlinear phase shifter ($k=2$), the QFI involves the variance
of the squared photon number:%
\begin{equation}
F_{2}=4\langle \Delta ^{2}\hat{n}_{a}^{2}\rangle .
\end{equation}%
where
\begin{equation}
\langle \Delta ^{2}\hat{n}_{a}^{2}\rangle =\left\langle \Psi _{\phi
}\right\vert \left( \hat{a}^{\dagger }\hat{a}\right) ^{4}\left\vert \Psi
_{\phi }\right\rangle -\left( \left\langle \Psi _{\phi }\right\vert \left(
\hat{a}^{\dagger }\hat{a}\right) ^{2}\left\vert \Psi _{\phi }\right\rangle
\right) ^{2}.
\end{equation}%
Using the normal ordering relations

\begin{eqnarray}
\left( \hat{a}^{\dagger }\hat{a}\right) ^{4} &=&\hat{a}^{\dagger 4}\hat{a}%
^{4}+6\hat{a}^{\dagger 3}\hat{a}^{3}+7\hat{a}^{\dagger 2}\hat{a}^{2}+\hat{a}%
^{\dagger }\hat{a},  \notag \\
\left( \hat{a}^{\dagger }\hat{a}\right) ^{2} &=&\hat{a}^{\dagger 2}\hat{a}%
^{2}+\hat{a}^{\dagger }\hat{a}.
\end{eqnarray}%
we derive

\begin{eqnarray}
F_{2} &=&4[\lambda ^{2}Q_{mn,4,4,0,0}+6\lambda ^{2}Q_{mn,3,3,0,0}+  \notag \\
&&7\lambda ^{2}Q_{mn,2,2,0,0}+\lambda ^{2}Q_{mn,1,1,0,0}-  \notag \\
&&(\lambda ^{2}Q_{mn,2,2,0,0}+\lambda ^{2}Q_{mn,1,1,0,0})^{2}]  \notag \\
&=&F_{1}+f.
\end{eqnarray}%
with
\begin{eqnarray}
f &=&4[\lambda ^{2}Q_{mn,4,4,0,0}+6\lambda ^{2}Q_{mn,3,3,0,0}+6\lambda
^{2}Q_{mn,2,2,0,0}  \notag \\
&&-(\lambda ^{2}Q_{mn,2,2,0,0}+\lambda ^{2}Q_{mn,1,1,0,0})^{2}  \notag \\
&&-(\lambda ^{2}Q_{mn,1,1,0,0})^{2}]
\end{eqnarray}%
where the strictly positive term $f$ involves higher-order moments of the
photon number distribution, namely $Q_{mn,4,4,0,0}$ and $Q_{mn,3,3,0,0}$.
Physically, this term captures the additional phase information encoded via
the nonlinear $\hat{n}_{a}^{2}$ dependence of the Kerr phase shifter. Such
an enhancement mechanism is absent in the linear phase shifter case.

\begin{figure}[tbh]
\label{Fig9} \centering%
\subfigure{
\begin{minipage}[b]{0.5\textwidth}
\includegraphics[width=0.83\textwidth]{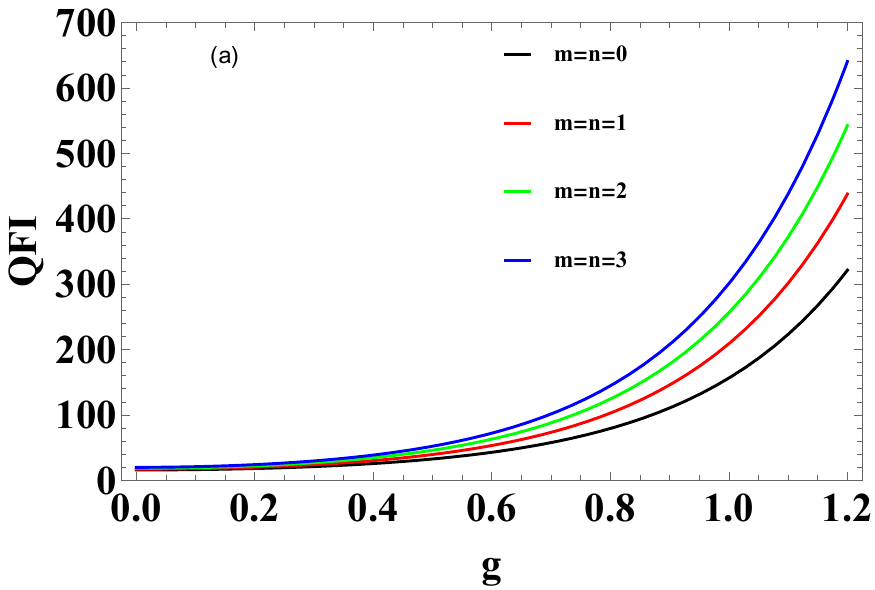}\\
\includegraphics[width=0.83\textwidth]{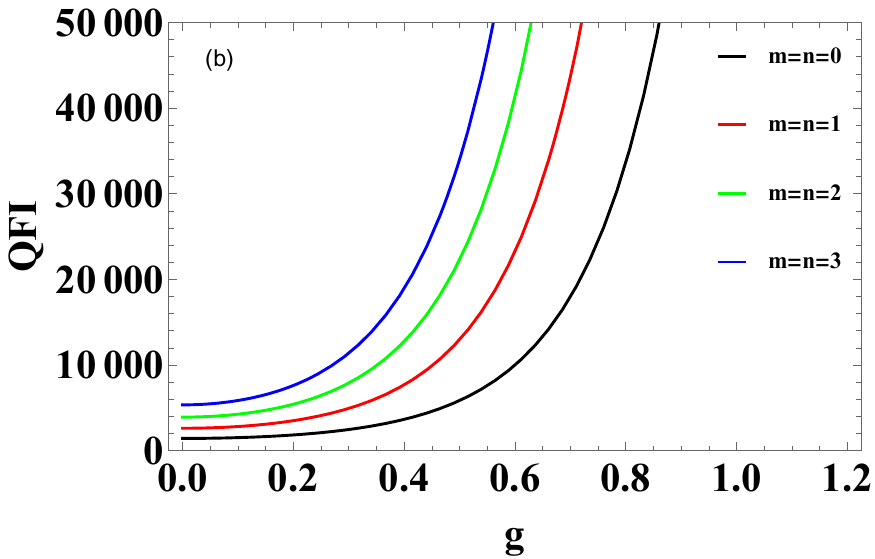}
\end{minipage}}
\caption{Quantum Fisher information (QFI) under ideal conditions as a
function of the OPA gain factor $g$ with $\protect\alpha =2$ for (a) the
linear phase shifter (Scheme I), and (b) the Kerr nonlinear phase shifter
(Scheme II).}
\end{figure}

A smaller $\Delta \phi _{QCRB}$, corresponding to a larger QFI, indicates
higher measurement sensitivity. Fig.~9 plots the QFI for both schemes as a
function of the OPA gain factor $g$ under MPS operations, with $\alpha =2$.
For both schemes, the QFI increases with increasing $g$. This enhancement is
further amplified as the photon-subtraction order increases. However, the
QFI values for Scheme II are dramatically larger than those for Scheme I
across the entire parameter range. This demonstrates the exceptional
effectiveness of the Kerr nonlinearity in elevating the ultimate information
content of the quantum state. The combination of Kerr nonlinearity and MPS
yields a QFI substantially higher than that of the linear counterpart.

\begin{figure}[tbh]
\label{Fig10} \centering%
\subfigure{
\begin{minipage}[b]{0.5\textwidth}
\includegraphics[width=0.83\textwidth]{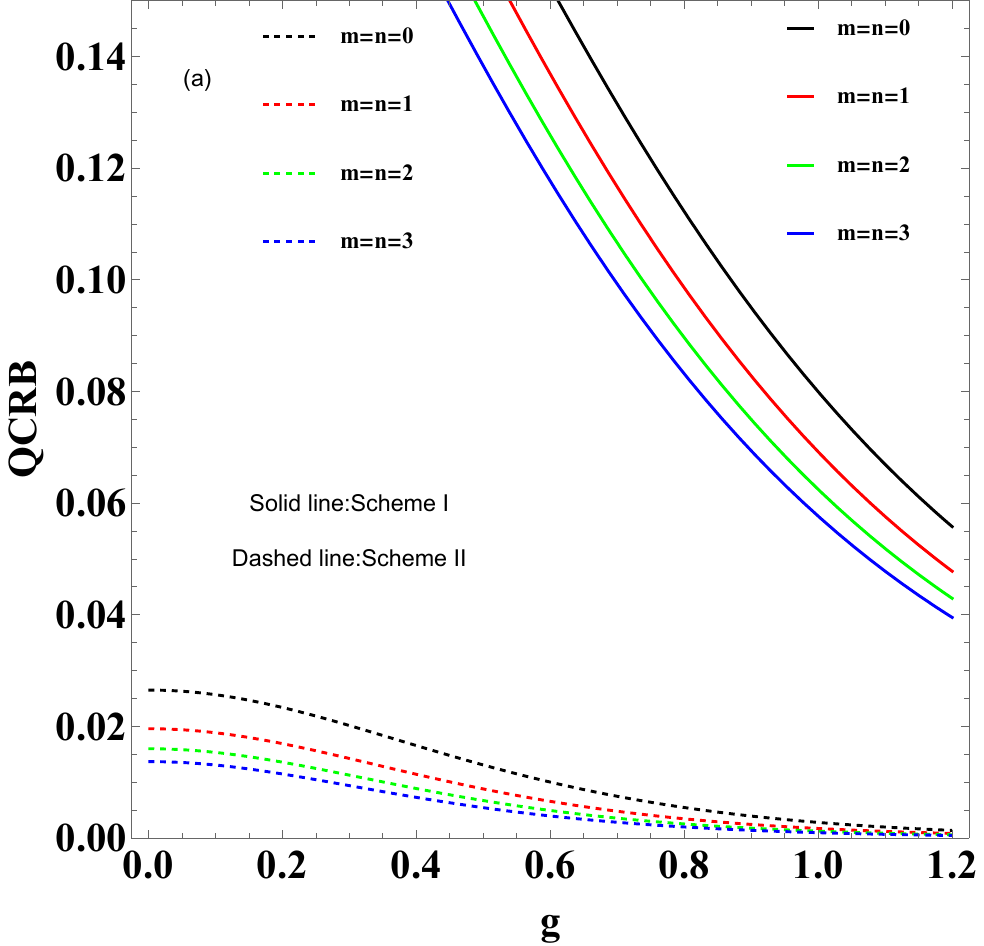}\\
\includegraphics[width=0.83\textwidth]{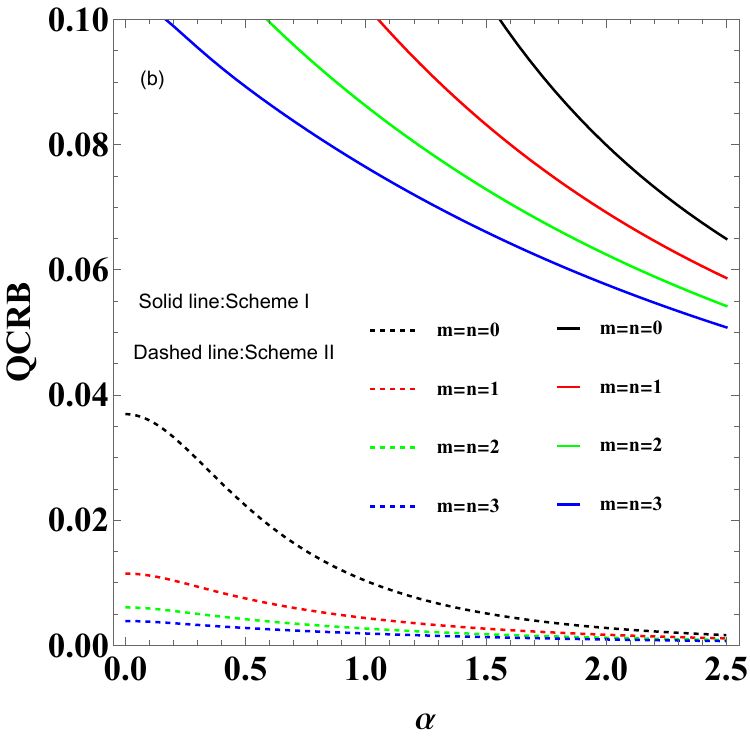}
\end{minipage}}
\caption{Quantum Cram\'{e}r-Rao bound (QCRB) under ideal conditions as
functions of (a) the OPA gain factor $g$ with $\protect\alpha =2$, and (b)
the coherent amplitude $\protect\alpha $ with $g=1$. Solid and dashed lines
correspond to Scheme I and Scheme II, respectively. The black, red, green,
and blue lines correspond to photon-subtraction orders $m=n=0$, $1$, $2$ and
$3$, respectively.}
\end{figure}

Fig.~10 presents the corresponding QCRB $\Delta \phi _{QCRB}$ as a function
of $g$ and $\alpha $. For both schemes, $\Delta \phi _{QCRB}$ decreases
significantly with increasing $g$ and $\alpha $. This confirms that greater
quantum resources yield a tighter ultimate precision bound. Crucially, for
identical parameters, the QCRB of Scheme II is consistently and
substantially lower than that of Scheme I. This superior performance
originates directly from the additional term $f$ in Eq.~(16). This term
captures the metrological gain provided by the nonlinear phase encoding
mechanism.

\subsection{The QCRB under Photon Loss}

For Scheme I, the derivation of the lossy QFI follows the general
purification-based framework of Escher et al. \cite{44} and the photon
subtraction approach detailed in Ref. \cite{30}. The loss channel is
simulated by coupling mode $a$ to an ancillary vacuum mode via a BS of
transmissivity $\eta ,$ yielding the expression in Eq. (18). For Scheme II,
the lossy QFI is obtained by extending the nonlinear Kerr phase estimation
framework developed in Ref. \cite{35}. By optimizing over a set of
variational parameters $\mu _{1}$ and $\mu _{2}$ that parametrize a general
measurement on the purified state, one obtains the tightest lower bound on
the QFI. Applying this procedure to the Kerr-encoded state $\left\vert \Psi
_{\phi }\right\rangle $ yields Eq. (19) with the coefficients $C_{i}$ given
in Appendix B.

For Scheme I, the QFI under loss ($F_{L1}$) is given by
\begin{equation}
F_{L1}=\frac{4F_{1}\eta \langle \hat{n}_{a}\rangle }{\left( 1-\eta \right)
F_{1}+4\eta \langle \hat{n}_{a}\rangle },
\end{equation}%
where $\eta $ denotes the transmittivity and $\left\langle \hat{n}%
_{a}\right\rangle =\lambda ^{2}Q_{mn,1,1,0,0}$ represents the mean photon
number in mode $a$ before loss.

For Scheme II, the lossy QFI ($F_{L2}$) takes a more complex form. This
complexity arises from the need to adapt the nonlinear phase measurement to
account for photon loss, and it is expressed as
\begin{eqnarray}
F_{L2} &=&4[C_{1}^{2}\langle \Delta ^{2}\hat{n}^{2}\rangle -C_{2}\langle
\hat{n}^{3}\rangle +C_{3}\langle \hat{n}^{2}\rangle -C_{4}\langle \hat{n}%
\rangle -  \notag \\
&&C_{5}\langle \hat{n}^{2}\rangle \langle \hat{n}\rangle -C_{6}\langle \hat{n%
}\rangle ^{2}].
\end{eqnarray}%
Here, the expectation values $\langle \cdot \rangle $ and variance $\langle
\Delta ^{2}\cdot \rangle $ are evaluated for the state $\left\vert \Psi
_{\phi }\right\rangle ,$ which corresponds to the state after the phase
shifter but before the BS. The coefficients $C_{i}$ (detailed in Appendix B)
depend on $\eta $ and two optimization parameters $\mu _{1opt}$, $\mu
_{2opt}.$ These parameters are introduced to adapt the measurement for
optimal extraction of phase information under loss, and their values are
obtained by maximizing $F_{L2}$ to yield the tightest precision bound. The
corresponding lossy QCRB is given by%
\begin{equation}
\Delta \phi _{QCRBL_{k}}=\frac{1}{\sqrt{F_{Lk}}}\quad (k=1,2).
\end{equation}

\begin{figure}[tbh]
\label{Fig11} \centering%
\subfigure{
\begin{minipage}[b]{0.5\textwidth}
\includegraphics[width=0.83\textwidth]{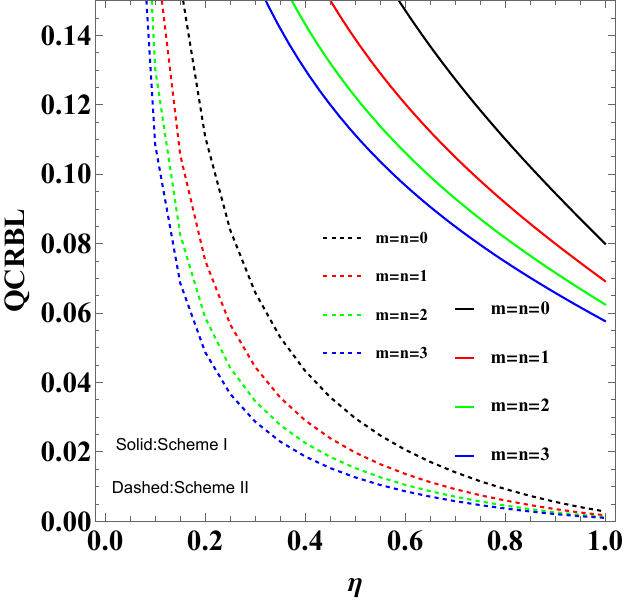}
\end{minipage}}
\caption{Lossy quantum Cram\'{e}r-Rao bound $\Delta \protect\phi %
_{QCRBL_{k}} $ as a function of transmissivity $\protect\eta $ with $\protect%
\alpha =2$ and $g=1$. Solid and dashed lines represent Scheme I and Scheme
II, respectively. Note that decreasing $\protect\eta $ corresponds to
increased photon loss.}
\end{figure}

Fig.~11 plots the lossy QCRB $\Delta \phi _{QCRBL_{k}}$ as a function of
transmissivity $\eta $. As physically expected, $\Delta \phi _{QCRBL_{k}}$
decreases with increasing $\eta ,$ which corresponds to decreasing internal
photon loss. The key result is that for all values of $\eta $, and
particularly in the high-loss regime where $\eta $ is small, $\Delta \phi
_{QCRBL_{2}}$ for Scheme II is vastly smaller than $\Delta \phi _{QCRBL_{1}}$
for Scheme I. Even in the absence of MPS and under moderate loss conditions,
Scheme II maintains a significantly lower ultimate precision bound. This
observation demonstrates that the Kerr nonlinear phase shifter not only
enhances the QCRB under ideal lossless conditions but also confers
remarkable robustness against internal photon loss, thereby preserving a
much tighter precision limit. The complexity of Eq.~(19) for $F_{L2}$, which
involves higher-order moments of the photon number distribution, enables the
system to partially retain the phase information encoded in these moments
despite photon loss. This is achieved through the optimized parameters $\mu
_{1opt}$ and $\mu _{2opt}$, which adapt the nonlinear measurement to
mitigate the deleterious effects of loss. This adaptive capability directly
explains the superior loss robustness observed in Fig.~11.

\section{Conclusion}

In summary, we have proposed and theoretically analyzed a high-precision
phase-estimation scheme that integrates a Kerr nonlinear phase shifter and
MPS operation within an HI architecture. Using a coherent state and a vacuum
state as inputs, we perform phase estimation via homodyne detection and
systematically evaluate the system performance under both ideal and lossy
conditions. We have carried out comprehensive investigations on phase
sensitivity, QFI, QCRB, and the influence of key parameters including the
coherent amplitude, OPA gain factor, and internal photon loss.

Our results reveal that the joint use of Kerr nonlinearity and MPS produces
strong synergistic effects and yields significant performance improvements
over schemes that rely on either technique alone. The nonlinear phase
shifter provides an intensity-dependent phase response proportional to the
squared photon number, which greatly enhances the signal dependence on phase
variations. MPS engineers non-Gaussian quantum states and further boosts
phase sensitivity. The proposed scheme can surpass the SQL, exceed the
conventional $1/N$ Heisenberg scaling, and approach the $1/N^{2}$\
super-Heisenberg scaling enabled by the Kerr nonlinearity. We reiterate that
this super-Heisenberg scaling is the intrinsic limit for the $k=2$ Kerr
nonlinearity and does not contradict the standard Heisenberg limit for
linear phase estimation. Quantitatively, for the same mean photon number $N$
and subtraction order, our Scheme II achieves a phase sensitivity
approximately two orders of magnitude lower than the HI with MPS in Ref.
\cite{18}, and surpasses the Kerr-enhanced SUI in Ref. \cite{35} by a factor
of about $3$ in the super-Heisenberg regime. This clearly demonstrates the
synergistic advantage of integrating Kerr nonlinearity and MPS within the
hybrid interferometer. In the presence of internal photon loss, the system
maintains excellent measurement precision and exhibits stronger robustness
than Scheme I. The QFI and QCRB analyses confirm that Scheme II achieves a
much tighter ultimate precision bound across a wide parameter range. The
enhanced robustness originates from the higher-order photon number moments
involved in the nonlinear QFI, which help preserve phase information even
under attenuation.

All components of the proposed scheme are experimentally feasible with
current quantum optical techniques. This work provides a practical and
effective strategy for constructing high-precision, noise-resilient quantum
metrology devices and promotes the development of real-world quantum sensors
and precision measurement applications.

\section{Acknowledgments}

This work is supported by the National Natural Science Foundation of China
(Grant No. 12564049 and Grant No. 12104195), the Jiangxi Provincial Natural
Science Foundation (Grants No. 20242BAB26009 and 20232BAB211033), as well as
the Jiangxi Provincial Key Laboratory of Advanced Electronic Materials and
Devices (Grant No. 2024SSY03011), Jiangxi Civil-Military Integration
Research Institute (Grant No. 2024JXRH0Y07), and the Science and Technology
Project of Jiangxi Provincial Department of Education (Grant No.
GJJ2404102). \newline
\newline
\textbf{Appendix A. Analytical Expressions for Phase Sensitivity} \newline

The expectation value of the operator $\hat{a}^{\dagger x1}\hat{a}^{y1}\hat{b%
}^{\dagger x2}\hat{b}^{y2}$ for the linear phase shifter (Scheme I, $k=1$)
is evaluated via a generating function method as
\begin{align}
& \lambda _{b}^{2}\left\langle 0\right\vert _{a}\left\langle \alpha
\right\vert \hat{U}_{OPA}^{\dagger }\hat{a}^{\dagger m}\hat{b}^{\dagger
n}\left( \hat{a}^{\dagger x1}\hat{a}^{y1}\hat{b}^{\dagger x2}\hat{b}%
^{y2}\right)  \notag \\
& \hat{a}^{m}\hat{b}^{n}\hat{U}_{OPA}\left\vert \alpha \right\rangle
_{a}\left\vert 0\right\rangle _{b}  \notag \\
& =\left\langle \hat{a}^{\dagger x1}\hat{a}^{y1}\hat{b}^{\dagger x2}\hat{b}%
^{y2}\right\rangle  \notag \\
& =\lambda ^{2}Q_{mn,x1,y1,x2,y2}.  \tag{A1}
\end{align}%
The general expression for phase sensitivity, as given by Eq.~(6), is:%
\begin{equation}
\Delta \phi _{k}=\frac{\sqrt{\left\langle \left( \hat{a}+\hat{a}^{\dag
}\right) ^{2}\right\rangle -\left\langle \hat{a}+\hat{a}^{\dag
}\right\rangle ^{2}}}{\left\vert \partial \left\langle \hat{a}+\hat{a}^{\dag
}\right\rangle /\partial \phi \right\vert }.  \tag{A2}
\end{equation}%
We provide analytical expressions for $\left\langle \hat{X}_{a}\right\rangle
$ and $\left\langle \hat{X}_{a}^{2}\right\rangle $ for both linear and Kerr
schemes using the generating function method. All expectation values are
derived consistently and evaluated under the same normalization condition.
For Scheme I, the specific expectation values required for Eq.~(A2) are
derived as follows:
\begin{align}
& \left\langle \psi _{out}\right\vert \hat{a}+\hat{a}^{\dag }\left\vert \psi
_{out}\right\rangle  \notag \\
& =\frac{\lambda ^{2}}{\sqrt{2}}(e^{-i\phi }Q_{mn,1,0,0,0}+iQ_{mn,0,0,1,0}+
\notag \\
& e^{i\phi }Q_{mn,0,1,0,0}-iQ_{mn,0,0,0,1}),  \tag{A3}
\end{align}%
and%
\begin{align}
& \left\langle \psi _{out}\right\vert \left( \hat{a}+\hat{a}^{\dag }\right)
^{2}\left\vert \psi _{out}\right\rangle  \notag \\
& =\lambda ^{2}(\frac{1}{2}e^{-2i\phi }Q_{mn,2,0,0,0}-\frac{1}{2}%
Q_{mn,0,0,2,0}+  \notag \\
& ie^{-i\phi }Q_{mn,1,0,1,0}+\frac{1}{2}e^{2i\phi }Q_{mn,0,2,0,0}-\frac{1}{2}%
Q_{mn,0,0,0,2}  \notag \\
& -ie^{i\phi }Q_{mn,0,1,0,1}+Q_{mn,1,1,0,0}-ie^{-i\phi }Q_{mn,1,0,0,1}+
\notag \\
& ie^{i\phi }Q_{mn,0,1,1,0}+Q_{mn,0,0,1,1}+Q_{mn,0,0,0,0}).  \tag{A4}
\end{align}

For the Kerr nonlinear phase shifter (Scheme II, $k=2$), the nonlinear phase
evolution $e^{2ti\phi (\hat{a}^{\dagger }\hat{a})}$ is handled by a more
involved generating function. The derivation proceeds by inserting coherent
state completeness relations after the Kerr unitary, converting the
expectation into multi-dimensional Gaussian integrals, and reorganizing the
result as a differential operator. The derivation proceeds as follows.
First, the transformation relations for the squeezer $\hat{U}_{OPA}^{\dagger
}\hat{a}^{\dagger }\hat{U}_{OPA},$the linear phase shifter Eq. (2), and the
Kerr nonlinearity Eq. (3) are applied sequentially to reduce the operator
products. Coherent state completeness relations $\frac{1}{\pi }\int
d^{2}z\left\vert z\right\rangle \left\langle z\right\vert $\ are then
inserted after the Kerr unitary to convert the resulting expectation value
into multi-dimensional Gaussian integrals over coherent state labels.
Evaluating these integrals in closed form and reorganizing the result as a
differential operator acting on a simpler exponential expression yields the
function $D_{mn,t,q,s}$\ defined below. We introduce:
\begin{align}
& \lambda _{b}^{2}\left\langle 0\right\vert _{a}\left\langle \alpha
\right\vert \hat{U}_{OPA}^{\dagger }\hat{a}^{\dagger m}\hat{b}^{\dagger
n}e^{2ti\phi (\hat{a}^{\dagger }\hat{a})}\hat{a}^{t}\hat{b}^{\dagger q}\hat{b%
}^{s}  \notag \\
& \hat{a}^{m}\hat{b}^{n}\hat{U}_{OPA}\left\vert \alpha \right\rangle
_{a}\left\vert 0\right\rangle _{b}  \notag \\
& =\lambda ^{2}D_{mn,t,q,s}e^{g5}.  \tag{A5}
\end{align}%
Here, $D_{mn,p,q,s}$ is a differential generating function defined as:

\begin{align}
& D_{mn,t,q,s}=\frac{\partial ^{2m+2n+t+q+s}}{\partial t_{1}^{m}\partial
\tau _{1}^{n}\partial t_{2}^{m}\partial \tau _{2}^{n}\partial
x_{1}^{t}\partial x_{2}^{q}\partial x_{3}^{s}}\times  \notag \\
& \frac{1}{1-\left( e^{2ti\phi }-1\right) \sinh ^{2}g}e^{g5}|_{t_{1}=\tau
_{1}=t_{2}=\tau _{2}=x_{1}=x_{2}=x_{3}=0}  \tag{A6}
\end{align}%
with the auxiliary functions:

\begin{align}
g1& =[t_{1}t_{2}+x_{1}t_{2}+\left( x_{2}+\tau _{2}\right) \left( x_{3}+\tau
_{1}\right) ]  \notag \\
& \times \sinh ^{2}g,  \notag \\
g2& =\alpha \cosh g\left( t_{1}+x_{1}\right) -\left\vert \alpha \right\vert
^{2},  \notag \\
g3& =\frac{\left( d_{1}+d_{1}\left( \left( e^{2ti\phi }-1\right) \cosh
g\right) \right) \left( d_{2}+d_{3}\right) }{1-\left( e^{2ti\phi }-1\right)
\sinh ^{2}g},  \notag \\
g4& =d_{3}\left( \frac{\alpha ^{\ast }}{\cosh g}+t_{1}+x_{1}\right)
+d_{4}\left( t_{2}\cosh g+\alpha \right) ,  \notag \\
g5& =g1+g2+g3+g4.  \tag{A7}
\end{align}%
The coefficients $d_{i}$ are defined as:%
\begin{align}
d_{1}& =\left( t1+x1\right) \left( e^{2ti\phi }-1\right) \sinh ^{2}g,  \notag
\\
d_{2}& =(t2\sinh ^{2}g+\alpha \cosh g),  \notag \\
d_{3}& =-\frac{1}{2}\left( x3+\tau _{1}\right) \sinh 2g,  \notag \\
d_{4}& =\left( \alpha ^{\ast }-\left( x2+\tau _{2}\right) \sinh g\right) .
\tag{A8}
\end{align}%
The expectation values for Scheme II are given by:
\begin{align}
& \left\langle \Psi _{out}\right\vert \left( \hat{a}^{\dagger }+\hat{a}%
\right) \left\vert \Psi _{out}\right\rangle  \notag \\
& =\frac{\lambda ^{2}}{\sqrt{2}}(e^{-i\phi }D_{mn,1,0,0}^{\dagger
}+iD_{mn,0,1,0}+  \notag \\
& e^{i\phi }D_{mn,1,0,0}+iD_{mn,0,0,1}),  \tag{A9}
\end{align}%
and%
\begin{align}
& \left\langle \Psi _{out}\right\vert \left( \hat{a}^{\dagger }+\hat{a}%
\right) ^{2}\left\vert \Psi _{out}\right\rangle  \notag \\
& =\lambda ^{2}(\frac{1}{2}e^{-4i\phi }D_{mn,2,0,0}^{\dagger }+ie^{-i\phi
}D_{mn,1,0,1}^{\dagger }  \notag \\
& -\frac{1}{2}D_{mn,0,2,0}+\frac{1}{2}e^{4i\phi }D_{mn,2,0,0}-ie^{i\phi
}D_{mn,1,0,1}  \notag \\
& -\frac{1}{2}D_{mn,0,0,2}+Q_{mn,1,1,0,0}-ie^{-i\phi }D_{mn,1,1,0}^{\dagger }
\notag \\
& +ie^{i\phi }D_{mn,1,1,0}+Q_{mn,0,0,1,1}+Q_{mn,0,0,0,0}).  \tag{A10}
\end{align}%
\newline
\textbf{Appendix B. Coefficients for the Lossy QFI of Scheme II} \newline
\newline
The coefficients $C_{i}$ are obtained by adapting the nonlinear phase
estimation framework of Ref. \cite{35} to the HI configuration, resulting in
the expressions below. The coefficients $C_{i}$ are given by:
\begin{align}
C_{1}& =c_{1}\eta ^{2}-2c_{2}\eta -\mu _{2},  \notag \\
C_{2}& =2\eta \left( 3c_{1}^{2}\eta ^{3}-3c_{3}\eta ^{2}-c_{4}\eta
+c_{5}\right) ,  \notag \\
C_{3}& =\eta \left( 11c_{1}^{2}\eta ^{3}-2c_{6}\eta ^{2}+c_{7}\eta
-4c_{1}c_{2}\right) ,  \notag \\
C_{4}& =\eta \left( 6\eta ^{3}-12\eta ^{2}+7\eta -1\right) c_{1}^{2},  \notag
\\
C_{5}& =2\eta \left( 1-\eta \right) c_{1}C_{1},  \notag \\
C_{6}& =\eta ^{2}\left( 1-\eta \right) ^{2}c_{1}^{2},  \tag{B1}
\end{align}%
where the intermediate parameters $c_{i}$ are%
\begin{align}
c_{1}& =1+2\mu _{1}-\mu _{2},  \notag \\
c_{2}& =\mu _{1}-\mu _{2},  \notag \\
c_{3}& =1+2\left( 3\mu _{1}-2\mu _{2}\right) +\left( 2\mu _{1}-\mu
_{2}\right) \left( 4\mu _{1}-3\mu _{2}\right) ,  \notag \\
c_{4}& =7\mu _{2}-6\mu _{1}+24\mu _{1}\mu _{2}-14\mu _{1}^{2}-9\mu _{2}^{2},
\notag \\
c_{5}& =\mu _{2}c_{1}-2c_{2}^{2},  \notag \\
c_{6}& =9+40\mu _{1}-22\mu _{2}+44\mu _{1}^{2}-48\mu _{1}\mu _{2}+13\mu
_{2}^{2},  \notag \\
c_{7}& =7+40\mu _{1}-26\mu _{2}+52\mu _{1}^{2}-64\mu _{1}\mu _{2}+19\mu
_{2}^{2}.  \tag{B2}
\end{align}%
The expression for $F_{L2}$ in Eq.~(19) is maximized with respect to two
free parameters $\mu _{1}$ and $\mu _{2}$. The optimal values $\mu _{1opt}$
and $\mu _{2opt}$ that achieve this maximization are given by:
\begin{align}
\mu _{1opt}& =\frac{be-cd}{ad-2\eta b^{2}},  \notag \\
\mu _{2opt}& =\frac{ae-2\eta bc}{ad-2\eta b^{2}},  \tag{B3}
\end{align}%
where we have defined the scalar quantities $a=2b_{1}H$, $b=b_{2}H$, $%
c=b_{3}H$, $d=b_{4}H$, and $e=\eta b_{5}H$. Here, $H$ is a column vector of
moments calculated from the state $\left\vert \Psi _{\phi }\right\rangle $:

\begin{equation}
H=\left( \left\langle \Delta ^{2}\hat{n}^{2}\right\rangle ,\left\langle \hat{%
n}^{3}\right\rangle ,\left\langle \hat{n}^{2}\right\rangle ,\left\langle
\hat{n}\right\rangle ,\left\langle \hat{n}^{2}\right\rangle \left\langle
\hat{n}\right\rangle ,\left\langle \hat{n}\right\rangle ^{2}\right) ^{T},
\tag{B4}
\end{equation}%
and row vectors $b_{j}\,(j=1,2,3,4,5)$ are defined using the auxiliary
parameters $a_{1}=\eta -1$, $a_{2}=6\eta ^{2}-6\eta +1$, $a_{3}=11\eta
^{2}-11\eta +2$, $a_{4}=2\eta -1$:%
\begin{align}
b_{1}& =\left( \eta a_{1},-a_{2},a_{3},-a_{2},2\eta a_{1},-\eta a_{1}\right)
,  \notag \\
b_{2}& =\left( a_{1}^{2},-3a_{1}a_{4},a_{3}-a_{4},-a_{2},a_{1}a_{4},-\eta
a_{1}\right) ,  \notag \\
b_{3}& =\left( \eta ^{2},-3\eta a_{4},a_{3}+a_{4},-a_{2},\eta a_{4},-\eta
a_{1}\right) ,  \notag \\
b_{4}& =\eta \left( \eta
^{-1}a_{1}^{3},-6a_{1}^{2},a_{3}-2a_{4},-a_{2},2a_{1}^{2},-\eta a_{1}\right)
,  \notag \\
b_{5}& =\left( \eta a_{1},-a_{2},a_{3},-a_{2},\eta ^{2}+a_{1}^{2},-\eta
a_{1}\right) .  \tag{B5}
\end{align}

\end{document}